\documentclass[letterpaper,aps,prd,twocolumn,tightenlines,preprintnumbers,nofootinbib,showkeys,superscriptaddress,longbibliography]{revtex4-1}
\pdfoutput=1

\usepackage[dvipsnames]{xcolor}
\usepackage{graphicx,epic,eepic,epsfig}
\usepackage[export]{adjustbox}
\usepackage{tikz,contour} 
\usetikzlibrary{shapes,arrows,positioning,automata,backgrounds,calc,er,patterns}
\usepackage[compat=1.1.0]{tikz-feynman}

\usepackage{fullpage}
\usepackage{relsize}
\usepackage[left=1.75cm,right=1.75cm,top=1.75cm,bottom=1.75cm]{geometry}

\usepackage{XCharter}
\usepackage{mathptmx}
\usepackage[T1]{fontenc}
\usepackage{pifont}
\usepackage{amsfonts,amsmath,amssymb,amsbsy,bm,mathtools,slashed,latexsym,esvect,simpler-wick}

\usepackage{float,multirow,array,cancel,dcolumn}
\usepackage{enumitem}
\usepackage{hyperref}
\hypersetup{colorlinks=true,citecolor=MidnightBlue,linkcolor=Brown,urlcolor=Brown}
\usepackage[caption=false,labelformat=simple]{subfig}
\usepackage{lipsum}

\usepackage{comment}
\usepackage{shorthand}
\usepackage{natbib}

\begin{document}

\title{Right-handed neutrino production through first-generation leptoquarks}

\author{Gokul Duraikandan}
\email{gokulduraikandan19@iisertvm.ac.in}
\affiliation{Indian Institute of Science Education and Research Thiruvananthapuram, Vithura, Kerala, 695 551, India}

\author{Rishabh Khanna}
\email{rishabh.khanna@research.iiit.ac.in}
\affiliation{Center for Computational Natural Sciences and Bioinformatics, International Institute of Information Technology, Hyderabad 500 032, India}

\author{Tanumoy Mandal}
\email{tanumoy@iisertvm.ac.in}
\affiliation{Indian Institute of Science Education and Research Thiruvananthapuram, Vithura, Kerala, 695 551, India}

\author{Subhadip Mitra}
\email{subhadip.mitra@iiit.ac.in}
\affiliation{Center for Computational Natural Sciences and Bioinformatics, International Institute of Information Technology, Hyderabad 500 032, India}
\affiliation{Center for Quantum Science and Technology, International Institute of Information Technology, Hyderabad 500 032, India}

\author{Rachit Sharma}
\email{rachit21@iisertvm.ac.in}
\affiliation{Indian Institute of Science Education and Research Thiruvananthapuram, Vithura, Kerala, 695 551, India}

\begin{abstract}\noindent
The collider phenomenology of leptoquarks (LQs) and right-handed neutrinos (RHNs) has been studied extensively in the literature. Because of the gauge singlet nature, the production of RHNs at the LHC is typically suppressed by the tiny light-heavy neutrino mixing angles. In this study, we explore a promising scenario where the presence of an LQ mediator significantly enhances RHN production. We focus on first-generation scalar and vector LQs interacting with the first-generation RHN. The prospects are better for the first-generation scenario than the other generations because of the enhanced parton distribution functions (PDFs) of first-generation quarks. The enhanced PDFs boost the production cross sections of LQs, particularly their single and indirect productions. Incorporating all production modes of LQs that result in a pair of RHNs, we estimate the discovery prospects by analysing the monoelectron and dielectron channels arising from the decay of the RHN pair. We find that the indirect production of LQs is crucial in determining the discovery reach at the HL-LHC for the first-generation scenario.
\end{abstract}

\maketitle 

\section{Introduction}
\noindent
Contemporary experimental advancements have established that at least two neutrino states possess tiny but nonzero masses ($\sim 0.1$ eV).
The type-I seesaw mechanism~\cite{Minkowski:1977sc, Mohapatra:1979ia} shows a simple way to generate the neutrino masses by introducing heavy right-handed neutrinos (RHNs) to the Standard Model (SM). However, the Majorana mass scale ($\Lambda_M$) of the RHNs must be very high---around the grand unification scale for Yukawa couplings of order unity---to make the light neutrinos sub-eV this way. Such a high mass scale renders the RHNs inaccessible at TeV-range colliders like the LHC. A Yukawa coupling of order $10^{-6}$ could bring $\Lambda_M$ down to the TeV range, but, in that case, the RHNs would be long-lived and likely to decay outside detectors. The inverse seesaw mechanism (ISM)~\cite{Mohapatra:1986aw, Mohapatra:1986bd} shows promise from the collider perspective, for it naturally contains TeV-scale RHNs that decay promptly, making them detectable at colliders.

Since RHNs are gauge singlets, the tiny light-heavy neutrino mixing angle severely suppresses their production at colliders. (Current constraints on the RHN mass and mixing angles are discussed in Ref.~\cite{Abdullahi:2022jlv}, while prospects for heavy RHN searches at future lepton colliders are discussed in Ref.~\cite{Banerjee:2015gca}.) An interesting possibility for RHN production arises from the decays of other beyond-the-SM particles, such as $W'$ bosons~\cite{Das:2017hmg,Keung:1983uu,ThomasArun:2021rwf}, $Z'$ bosons~\cite{Ekstedt:2016wyi,Das:2017flq,Das:2017deo,Cox:2017eme,Das:2018tbd,Choudhury:2020cpm,Deka:2021koh,Das:2022rbl,Arun:2022ecj}, or leptoquarks (LQs or $\ell_q$)~\cite{Evans:2015ita,Bhaskar:2023xkm}, etc. Since, in these cases, the production is no longer dependent on the active-sterile mixing, they can be produced easily. From the LHC perspective, the production via TeV-scale LQs looks particularly promising~\cite{Bhaskar:2023xkm}. 

LQs are hypothetical coloured bosons (scalar or vector) carrying both baryon and lepton numbers. Consequently, they bridge the baryon and lepton sectors, enabling couplings between quarks and the RHNs. LQs naturally emerge in many extensions of the SM that aim to unify matter or forces. They are integral to a wide range of theoretical frameworks beyond the SM, such as the Pati-Salam model~\cite{Pati:1973uk,Pati:1974yy}, Grand Unified Theories~\cite{Georgi:1974sy,Fritzsch:1974nn}, composite models~\cite{Schrempp:1984nj}, coloured Zee-Babu models~\cite{Kohda:2012sr}, technicolor models~\cite{Dimopoulos:1979es,Farhi:1980xs} and Supersymmetry with $R$-parity violation~\cite{Barbier:2004ez}, etc. Recently, LQs have garnered significant attention as potential explanations for various low-energy anomalies (see, e.g., Refs.~\cite{Aydemir:2022lrq,Bhaskar:2022vgk,Bhaskar:2024swq}). Their collider phenomenology is also well-studied in the literature (e.g., Refs.~\cite{Mandal:2015vfa,Bhaskar:2020kdr, Bandyopadhyay:2021pld, Cheung:2023gwm,Bhaskar:2024snl,Bhaskar:2024wic}).

Previously, we studied the production of RHNs through second-generation LQs at the LHC~\cite{Bhaskar:2023xkm}. Here, we focus on RHN productions via first-generation LQs. These LQs couple to a first-generation quark and the first-generation RHN, whose decay produces a first-generation lepton in the final state. In such a scenario, RHN pairs can be produced at the LHC through all LQ production mechanisms~\cite{Bhaskar:2023xkm}: pair and single productions (P- and SPs), where the produced LQs decay to RHNs, and the indirect production (IP)~\cite{Bhaskar:2023ftn} of a LQ, where a $t$-channel LQ exchange produces two RHNs ($pp\to\n_R\n_R$). For our collider study, we assume that a TeV-range first-generation LQ decays exclusively to a lighter first-generation RHN at the tree level, i.e., the branching ratio (BR) of the $\ell_q \to q \nu_R$ decay is essentially $100$\%.  The nature of the LQ couplings strongly influences the SP and IP cross sections, as the quark parton distribution functions (PDFs) vary significantly across generations. As a result, the first-generation case requires a separate study from the second-generation one, even though the detection efficiencies of the first and second-generation leptons (i.e., $e$ and $\m$) are not significantly different. However, the third-generation case differs from those two both in terms of initial-quark PDFs and final-state isolation/detection strategies. We plan to address the third-generation case in a separate study.

\begin{table*}[]
\caption{List of LQs and their representations under the SM gauge groups. In the second column, we show the representations under $SU(3)_c$, $SU(2)_L$, and $U(1)_Y$, respectively. In the last column, we present
the LQ interactions with $\nu_R$~\cite{Dorsner:2016wpm}. 
\label{tab:1}}
\centering
{
\renewcommand\baselinestretch{2}\selectfont
\begin{tabular*}{\textwidth}{l@{\extracolsep{\fill}} ccc}
\hline
    LQ & ($SU(\mathbf{3})_{C},\ SU(2)_{L},\ U(1)_{Y}$) & Spin & Interaction Lagrangian\\
    \hline \hline
    ${S}_1$ & $(\mathbf{\Bar{3}},1,1/3)$& $0$ & ${y}_{1ij}^{LL} \overline{Q^c_{L}}^{i,a}  {S}_{1}  \varepsilon^{ab} L_{L}^{j,b} + {y}_{1ij}^{RR} \overline{u^{c}}_{R}^{i} {S}_{1} {e}_{R}^{j} + \overline{y}_{1ij}^{{RR}} \overline{d^{c}}_{R}^{i} {S}_{1} \nu_{R}^{j}$ + h.c. \\
    $\overline{S}_{1}$ & $(\mathbf{\Bar{3}},1,-2/3)$ & $0$ & $\overline{y}_{1ij}^{{RR}} \overline{u^{c}}_{R}^{i}  \overline{S}_{1} \nu_{R}^{j}$  + h.c. \\ 
    $\widetilde{R}_{2}$ & $(\mathbf{3},2,1/6)$ & $0$ &$- \widetilde{y}_{2ij}^{RL} \overline{d}_{R}^{i}  \widetilde{R}_{2}^{a}  \varepsilon^{ab} L_{L}^{j,b} + \widetilde{y}_{2ij}^{LR} \overline{Q}_{L}^{i,a} \widetilde{R}_{2}^{a}  \nu_{R}^{j}$ + h.c.  \\
    ${U}_1$ & $(\mathbf{3},1,2/3)$& $1$ & $ {x}_{1ij}^{LL} \overline{Q}_{L}^{i,a} \gamma^{\mu} {U}_{1,\mu} L_{L}^{j,a} + {x}_{1ij}^{RR} \overline{d}_{R}^{i} \gamma^{\mu} {U}_{1,\mu} {e}_{R}^{j} + {x}_{1ij}^{RR} \overline{u}_{R}^{i} \gamma^{\mu} {U}_{1,\mu} \nu_{R}^{j}$ + h.c. \\
    $\overline{U}_{1}$ & $(\mathbf{3},2,-1/3)$ & $1$ & $\overline{x}_{1ij}^{RR} \overline{d}_{R}^{i} \gamma^{\mu} \overline{U}_{1,\mu} \nu_{R}^{j}$  + h.c. \\
    $\widetilde{V}_{2}$ & $(\mathbf{\Bar{3}},1,-1/6) $& $1$ & $\widetilde{x}_{2ij}^{RL} \overline{u^{c}}_{R}^{i} \gamma^{\mu} \widetilde{V}_{2,\mu}^{b}  \varepsilon^{ab} L_{L}^{j,a} + \widetilde{x}_{2ij}^{LR} \overline{Q^{c}}_{L}^{i,a} \gamma^{\mu} \varepsilon^{ab} \widetilde{V}_{2,\mu}^{b}  \nu_{R}^{j}$ + h.c. \\
    \hline
    \end{tabular*}
}
\end{table*}

If a LQ decays predominantly to the RHN, it is not bound by any direct or indirect collider constraints---the LHC has yet to probe and constrain this part of the LQ parameter space. The RHN production through LQ decay becomes significant when neither of these particles is too heavy, and the RHN is lighter than the LQ. Since LQs are strongly interacting particles, they can be abundantly produced at the LHC, enhancing the RHN production. This process depends on a few parameters, such as the RHN and LQ masses, the RHN-LQ-quark couplings, and the PDFs of the initial partons. The cross section of the $qq\to\n_R\n_R$ process through a $t$-channel LQ exchange is proportional to the fourth power of the RHN-LQ-quark couplings. Hence, this coupling can not be small for the LQ-exchange process to be significant. 

In this paper, we analyse RHN production through all the above-mentioned channels at the LHC. The plan of the paper is as follows. In the next section, we list the LQ models with RHN decay mode. In Section~\ref{sec:LHCpheno}, we discuss various possible final states arising from the decay of a pair of RHN that can be searched for at the High Luminosity-LHC (HL-LHC). We also discuss the SM backgrounds and selection criteria for the monoelectron and dielectron channels. We present our findings in Section~\ref{sec:HLLHCpros} and conclude in Section~\ref{sec:conclu}.

\section{Leptoquark models}
\noindent
We consider all possible LQs that simultaneously interact with a first-generation quark and a RHN. Table~\ref{tab:1} lists the scalar and vector LQs (sLQs and vLQs) with the necessary couplings. Since we are only interested in the collider phenomenology of these models, we ignore the diquark operators to bypass the proton-decay restrictions. Following Ref.~\cite{Dorsner:2016wpm}, we denote the Yukawa coupling matrices of sLQs and vLQs with quark-lepton pairs generically as \( y \) and \( x \), respectively.  The first subscript of the couplings indicates the \( SU(2)_L \) representation and the superscripts indicate the chiralities of the fermions (the first one denotes the quark chirality and the second one, the lepton chirality). For simplicity, we assume both the Pontecorvo-Maki-Nakagawa-Sakata (PMNS) neutrino-mixing and the Cabibbo-Kobayashi-Maskawa (CKM) quark-mixing matrices to be approximately identity. This is justified since, unlike the second-generation case~\cite{Bhaskar:2023xkm}, the cross sections of the processes initiated by the first-generation quarks are not affected significantly by the presence of the small off-diagonal terms in the quark-mixing matrix, and the LHC experiments cannot identify the flavour of the missing neutrinos. Vector LQs can have an extra gluon coupling allowed by the gauge symmetries that can affect their productions at the LHC~\cite{Blumlein:1996qp,Blumlein:1994tu}:
\begin{align}
\mathcal{L} \supset & 
 - i g_{s} (1-\kappa) \chi^{\dagger}_{\mu} T^{a} \chi_{\nu} G^{a \mu\nu},
\end{align}
where $\chi_{\mu}$ denotes a generic vLQ and $G^{a\mu\nu}$ denotes the field strength tensor of the gluon. 

\section{LHC Phenomenology}
\label{sec:LHCpheno}

\noindent
We incorporate the new physics Lagrangian in \textsc{FeynRules}~\cite{Alloul:2013bka} to produce the model files for further simulation. We use \textsc{MadGraph5}~\cite{Alwall:2014hca} to generate the leading-order signal and background events with the \textsc{NNPDF23LO1}~\cite{NNPDF:2021uiq} PDF set using the default dynamical scale setting. Where available, we use appropriate $K_{\rm QCD}$ factors to account for higher-order cross sections. In the signal, we use an average NLO $K_{\rm QCD}$ factor of $1.58$~\cite{Kramer:2004df,Mandal:2015lca,Borschensky:2020hot,Borschensky:2021hbo,Borschensky:2021jyk,Borschensky:2022xsa} for the sLQ pair production. The events are passed through \textsc{Pythia8}~\cite{Bierlich:2022pfr} (for hadronization and showering) and \textsc{Delphes3}~\cite{deFavereau:2013fsa} (for detector simulation using the standard CMS card). We use \textsc{FastJet}'s anti-clustering algorithm~\cite{Cacciari:2011ma,Cacciari:2008gp} to reconstruct the jets from the \textsc{Delphes} tower objects. In our study, we use jets with two radii: (a) jets with $R=0.4$ (we call this an AK4 jet and denote it as $j$) and (b) fatjet with $R =1.2$ (denoted by $J$). We denote charged leptons with $\ell = \{e,\mu,\ta\}$.

\begin{figure}[t!]
  \centering
  \includegraphics[width=\columnwidth]{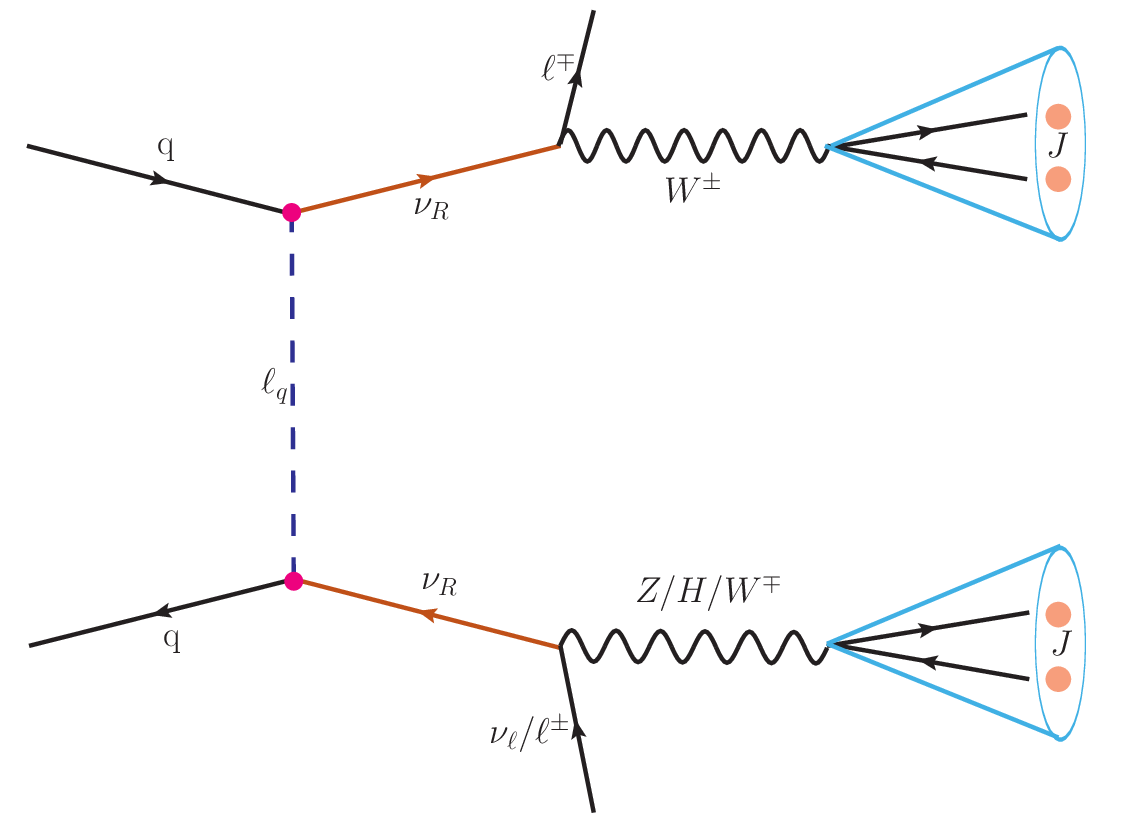}
  \caption{Mono- and di-lepton final states from RHN pair production through a $t$-channel LQ exchange at the LHC.}
  \label{fig:Feyn}
\end{figure}

\begin{figure*}
    \centering
    \captionsetup[subfigure]{labelformat=empty}
    \subfloat[\quad\quad\quad(a)]{\includegraphics[width=0.35\textwidth]{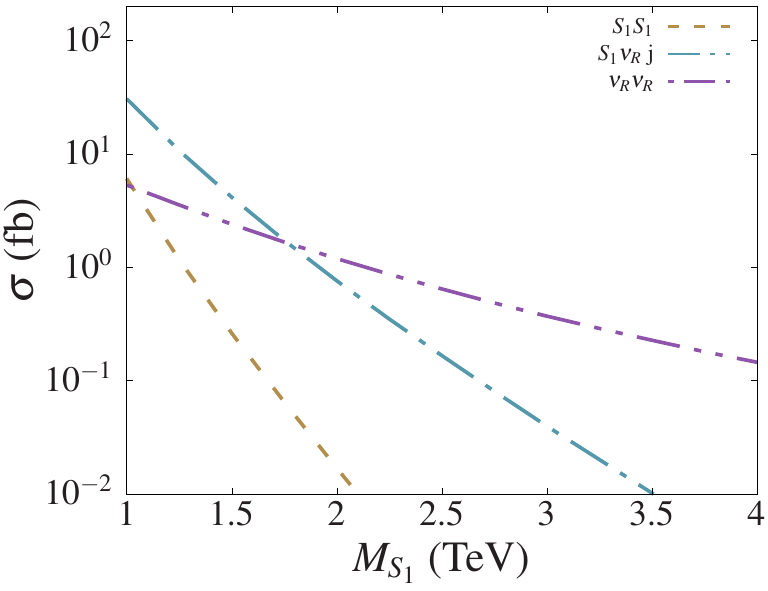}\label{S1_CS}}\hspace{1cm}
    \subfloat[\quad\quad\quad(b)]{\includegraphics[width=0.35\textwidth]{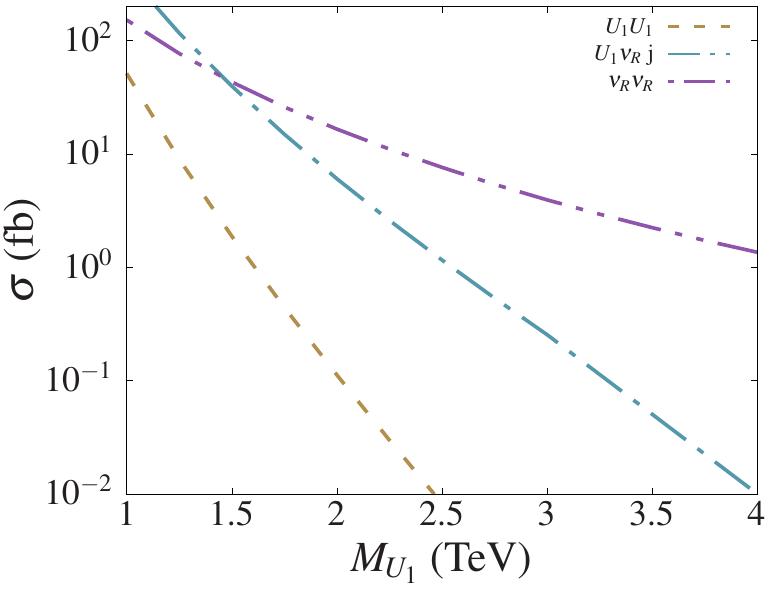}\label{U1_CS}}\\
    \subfloat[\quad\quad\quad(c)]{\includegraphics[width=0.35\textwidth]{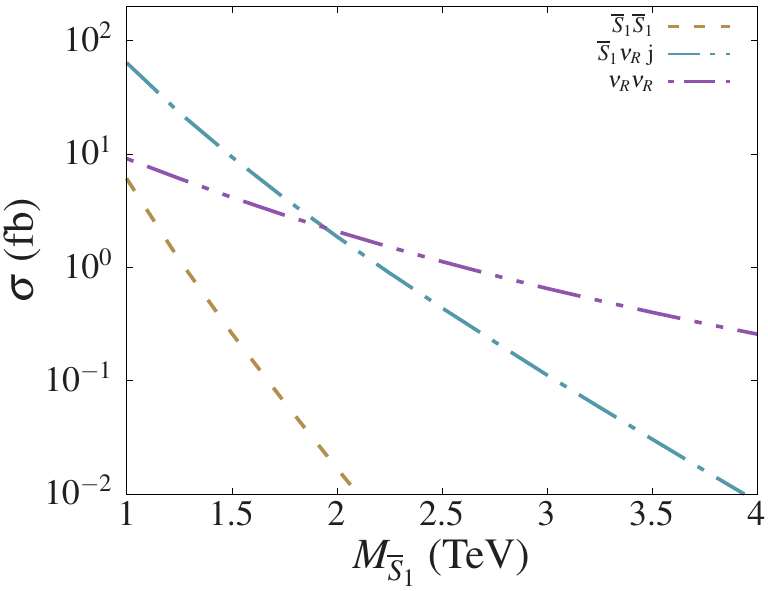}\label{S1bar_CS}}\hspace{1cm}
    \subfloat[\quad\quad\quad(d)]{\includegraphics[width=0.35\textwidth]{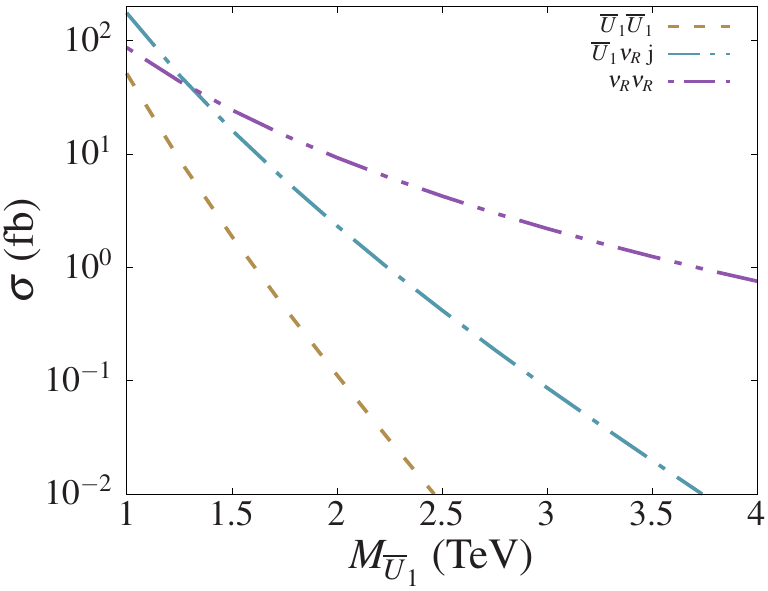}\label{U1bar_CS}}    
    \caption{Cross sections of direct and indirect production modes of charge-\(2/3\) and \(1/3\) sLQs (left) and vLQs (right) at the $14$~TeV LHC. For these plots, we set $M_{\nu_R}=500$~GeV and $x,y=1$. The vLQ plots are obtained for $\kappa =1$. \label{MLQvsCS}}
\end{figure*}

\begin{table}[!b]
\caption{Higher-order cross sections of the major background processes without decay and any cut. Their QCD orders are shown in the last column. We use these cross sections to compute the $K$ factors to incorporate the higher-order effects.\label{tab:Backgrounds}}
\centering{\small\renewcommand\baselinestretch{1.35}\selectfont
\begin{tabular*}{\columnwidth}{l @{\extracolsep{\fill}} crc }
\hline
\multicolumn{2}{l}{Background } & $\sg$ & QCD\\ 
\multicolumn{2}{l}{processes}&(pb)&order\\\hline\hline
\multirow{2}{*}{$V +$ jets~ \cite{Catani:2009sm,Balossini:2009sa}  } & $Z +$ jets  &  $6.33 \times 10^4$& N$^2$LO \\ 
                & $W +$ jets  & $1.95 \times 10^5$& NLO \\ \hline
\multirow{3}{*}{$VV +$ jets~\cite{Campbell:2011bn}}   & $WW +$ jets  & $124.31$& NLO\\ 
                  & $WZ +$ jets  & $51.82$ & NLO\\ 
                   & $ZZ +$ jets  &  $17.72$ & NLO\\ \hline
\multirow{3}{*}{Single $t$~\cite{Kidonakis:2015nna}}  & $tW$  &  $83.10$ & N$^2$LO \\ 
                   & $tb$  & $248.00$ & N$^2$LO\\ 
                   & $tj$  &  $12.35$ & N$^2$LO\\  \hline
$tt$~\cite{Muselli:2015kba}  & $tt +$ jets  & $988.57$ & N$^3$LO\\ \hline
\multirow{2}{*}{$ttV$~\cite{Kulesza:2018tqz}} & $ttZ$  &  $1.05$ &NLO+N$^2$LL \\
                   & $ttW$  & $0.65$& NLO+N$^2$LL \\ \hline
\end{tabular*}}
\end{table}
\begin{table*}
\caption{Selection cuts applied on the monoelectron and dielectron final states. \label{tab:Cuts}}
\centering
\renewcommand\baselinestretch{1.5}\selectfont
\begin{tabular*}{\textwidth}{@{\extracolsep{\fill}} lll}
\hline
\multicolumn{1}{r}{\multirow{1.5}{*}{Selection Cuts}} & \multicolumn{2}{c}{Channels}                   \\ \cline{2-3} 
\multicolumn{1}{r}{}                                & Monoelectron         & Dielectron              \\ \hline \hline
C1                                                 & \begin{tabular}[c]{@{}l@{}}$p_{T}(e) > 250$ GeV \\ $p_{T}(j_1),\ p_{T}(j_2) > 110$ GeV\end{tabular}   & \begin{tabular}[c]{@{}l@{}}$p_{T}(e) > 120$ GeV \\ No $b$-tagged jet \end{tabular} \\ \hline
C2   & \begin{tabular}[c]{@{}l@{}}$p_{T}(J_1) > 350$ GeV,  \\  $\eta ({J}_1) < 2.5$ \\ $0.03 < \tau_{21}(J_1) < 0.4$ \end{tabular} & \begin{tabular}[c]{@{}l@{}}$p_{T}(J_1) > 280$ GeV,  \\  $\eta ({J}_1) < 2.5$ \end{tabular}  \\  \hline
C3                                                  &                   & \begin{tabular}[c]{@{}l@{}}$M(e_1,e_2) > 250$ GeV  \\  $M(J_{1},e_1) > 450$ GeV\end{tabular}                  \\ \hline
C4                                                & \begin{tabular}[c]{@{}l@{}}$\slashed{E}_T >  180$ GeV\\ $S_T >  900$ GeV\end{tabular} & $S_T >  1200$ GeV\\ \hline
\end{tabular*}
\end{table*}
\begin{figure*}[]
    \centering
    \captionsetup[subfigure]{labelformat=empty}
    \subfloat[(a)]{\includegraphics[width=0.25\textwidth]{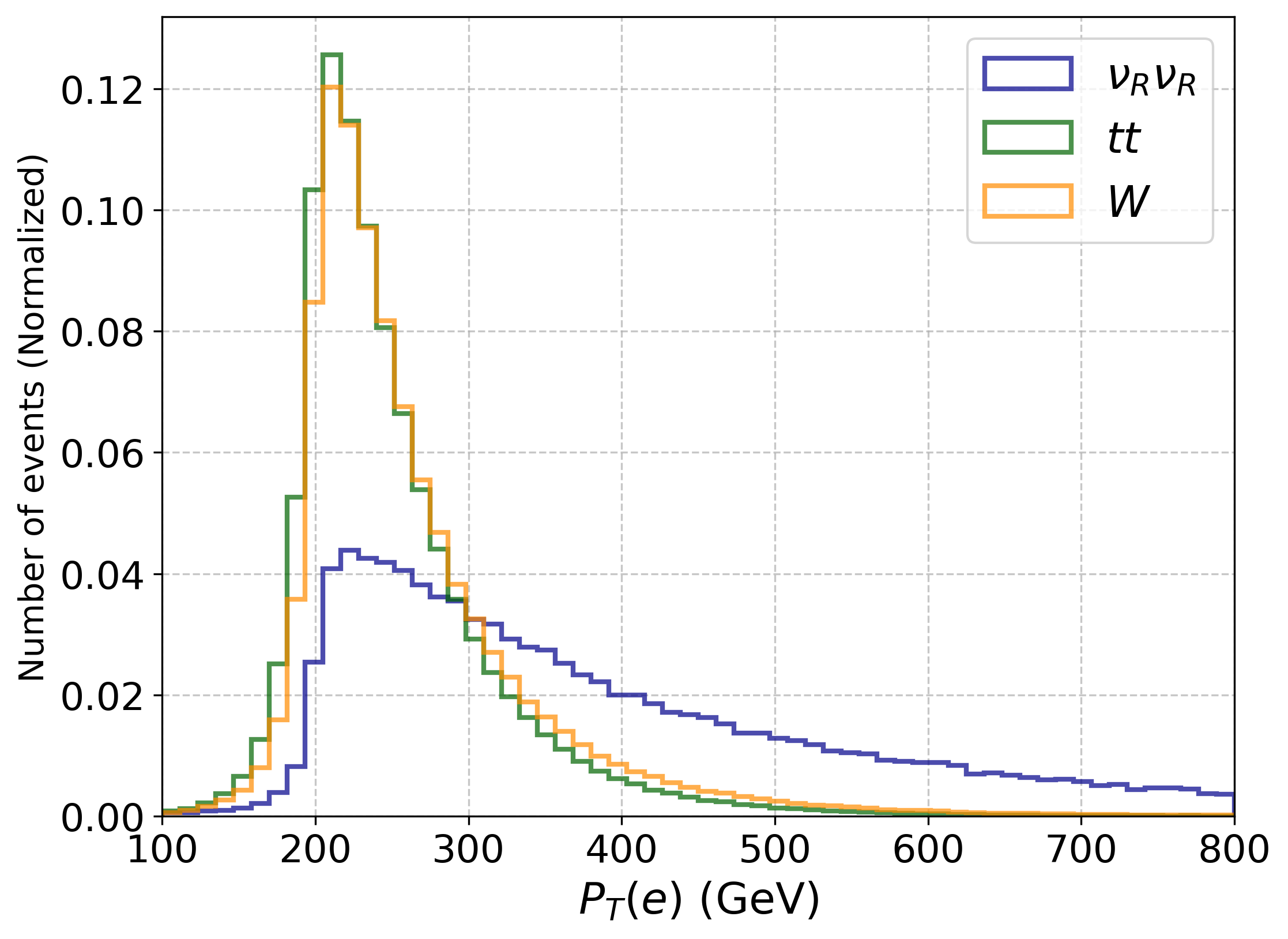}\label{pt_e_mono}}\hfill
    \subfloat[(b)]{\includegraphics[width=0.25\textwidth]{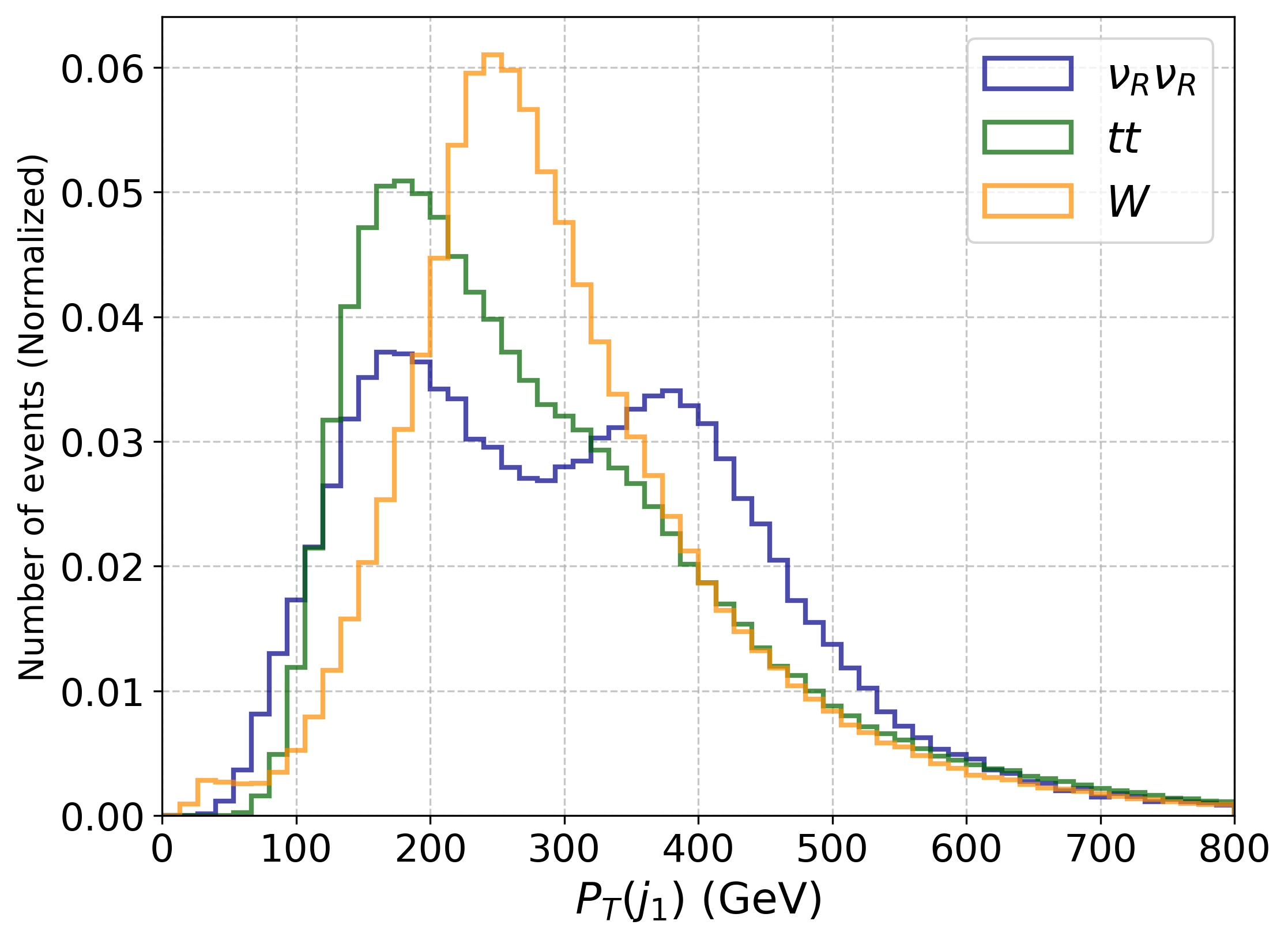}\label{pt_j_mono}}\hfill
    \subfloat[(c)]{\includegraphics[width=0.25\textwidth]{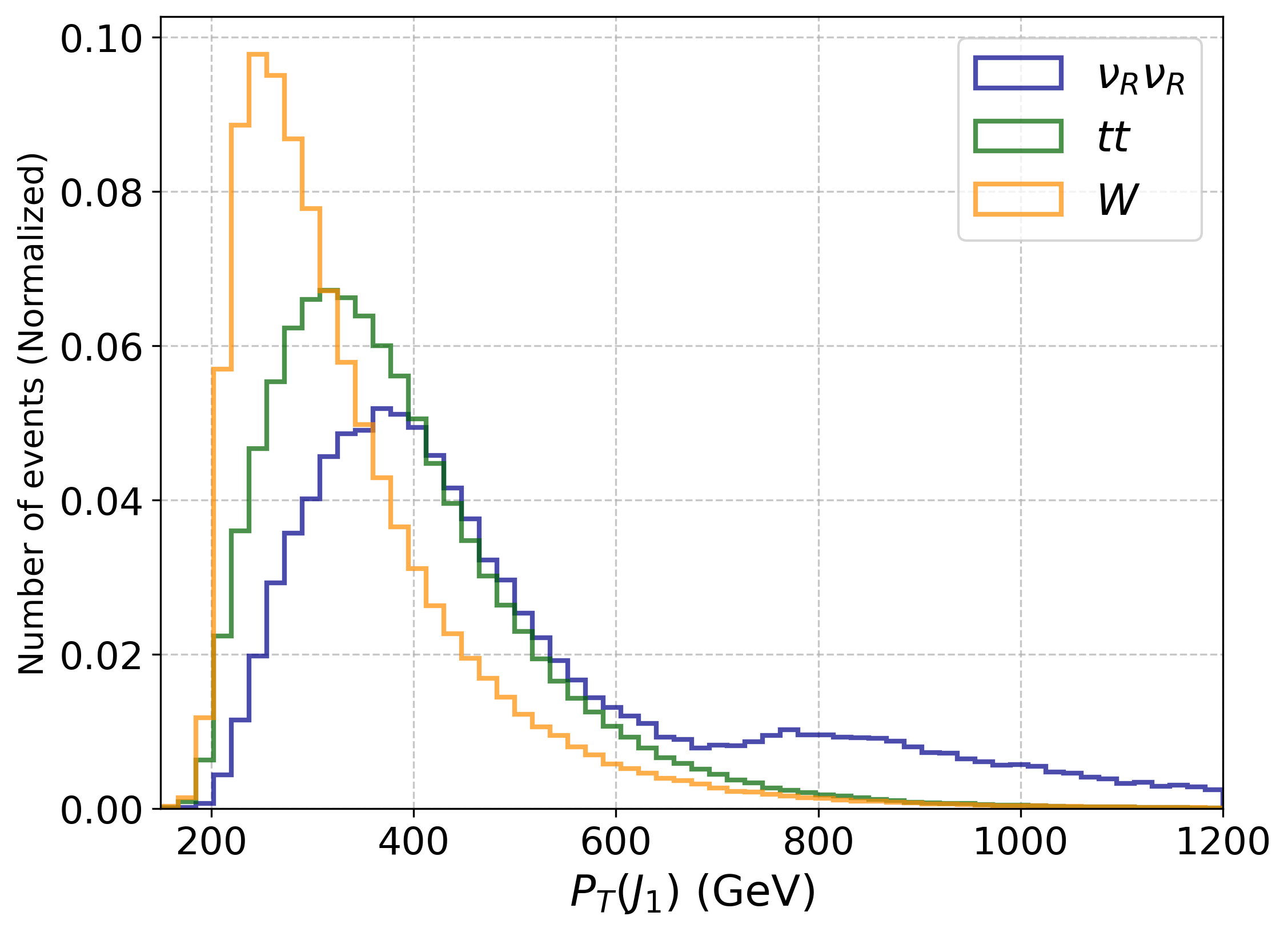}\label{pt_fjt_mono}}\hfill
    \subfloat[(d)]{\includegraphics[width=0.25\textwidth]{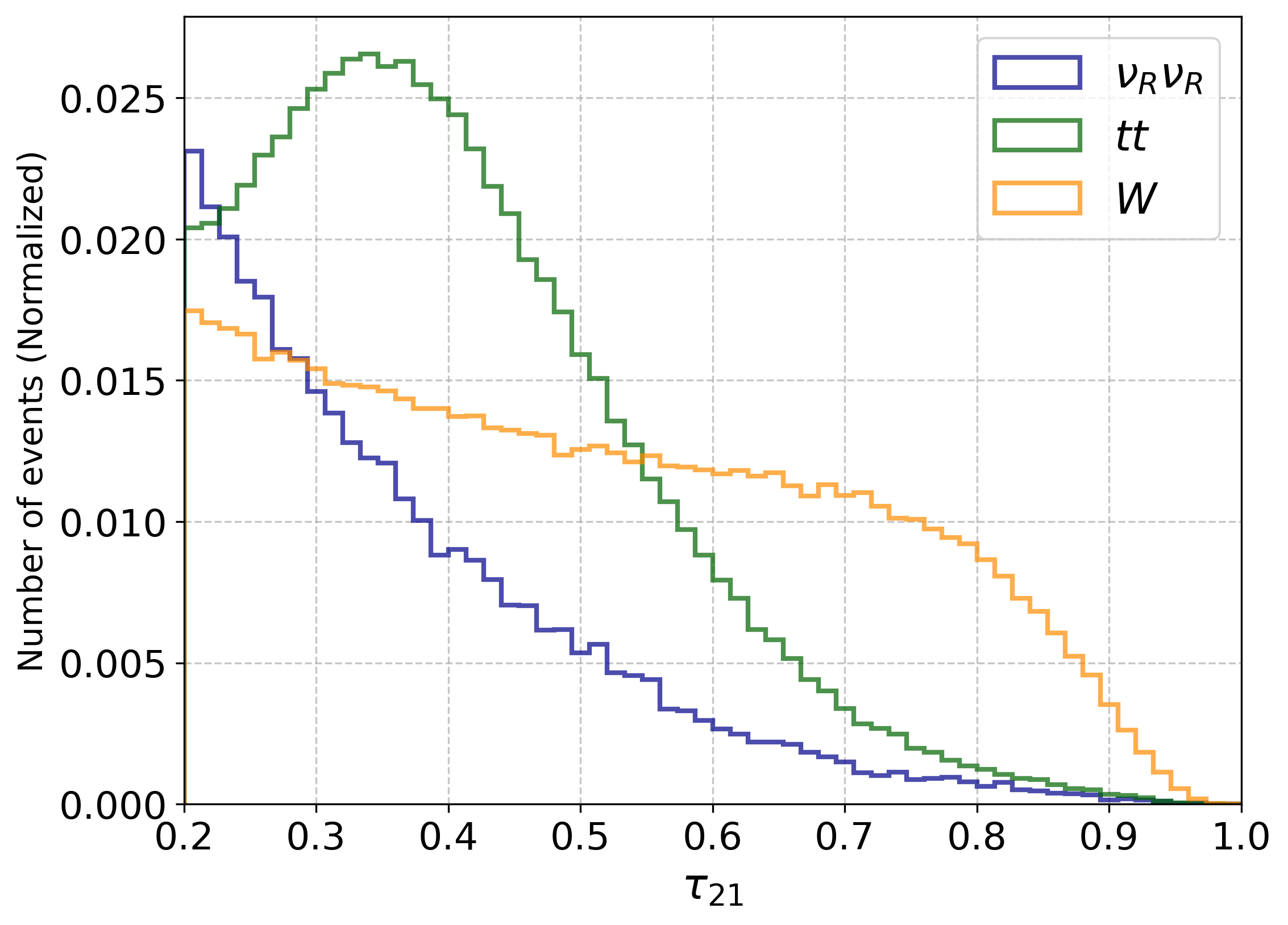}\label{tau21_mono}}\\
    \subfloat[(e)]{\includegraphics[width=0.25\textwidth]{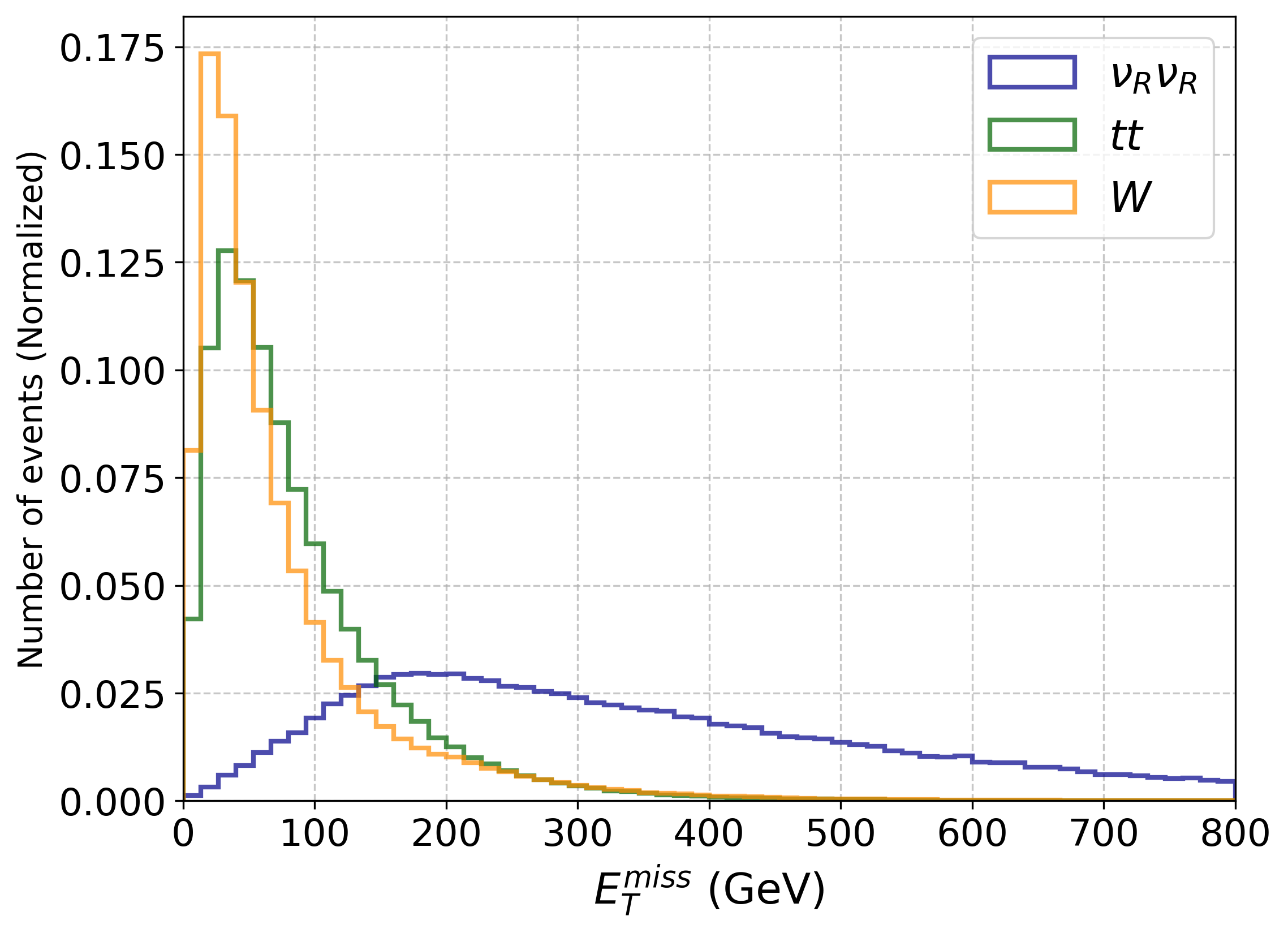}\label{met_mono}}\hfill
    \subfloat[(f)]{\includegraphics[width=0.25\textwidth]{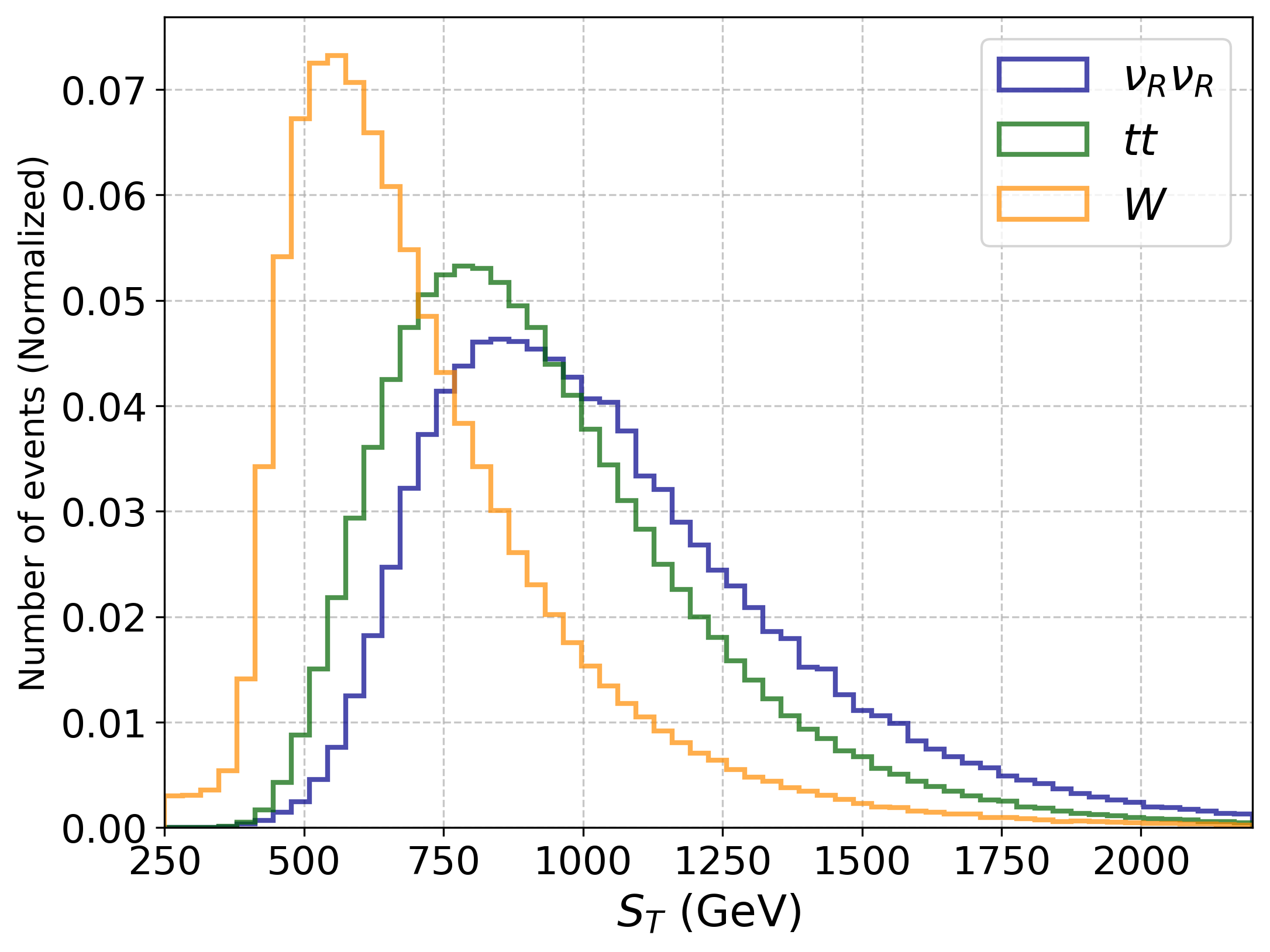}\label{ST_mono}}\hfill
    \subfloat[(g)]{\includegraphics[width=0.25\textwidth]{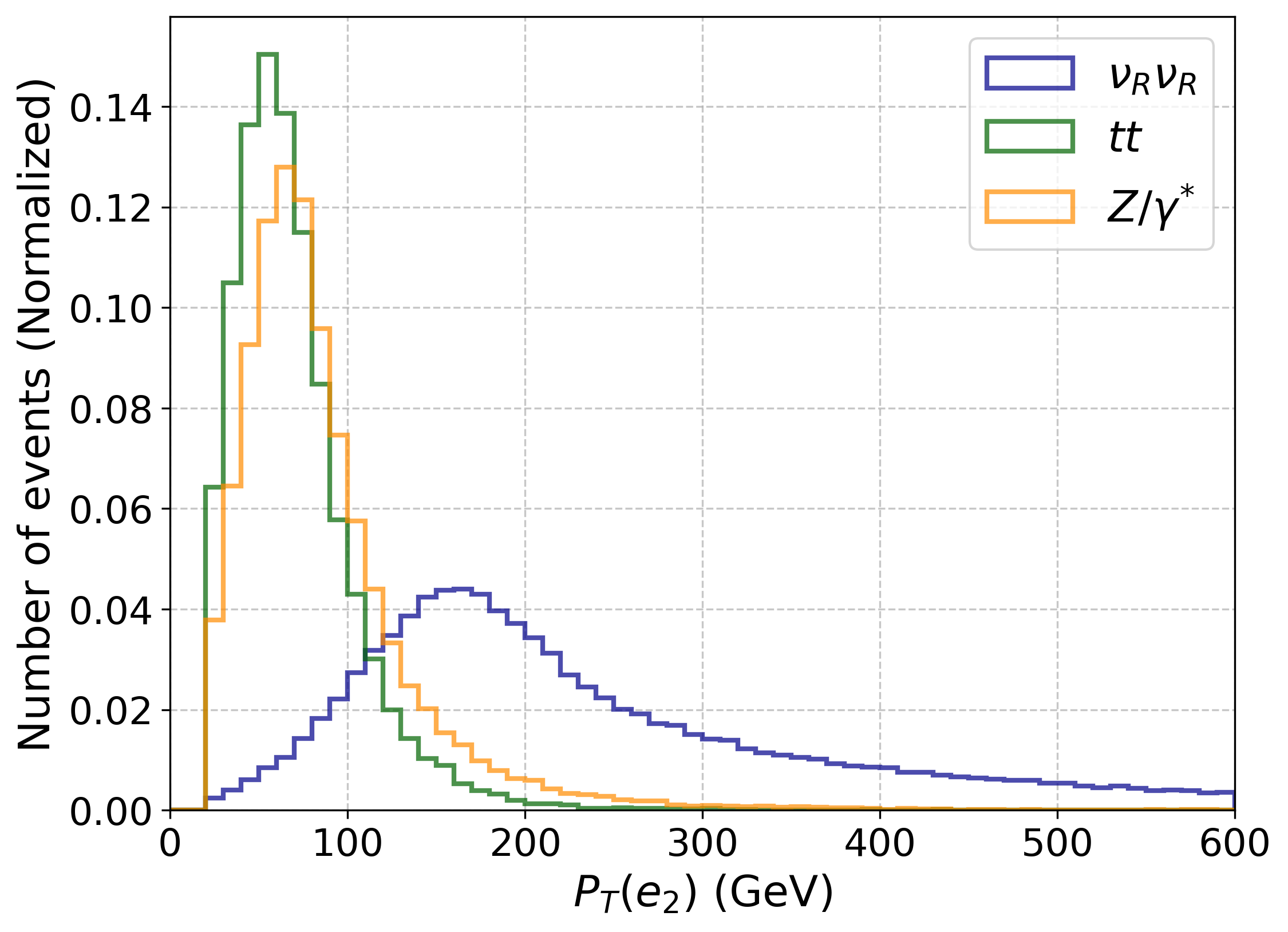}\label{pt_e_dilep}}\hfill
    \subfloat[(h)]{\includegraphics[width=0.25\textwidth]{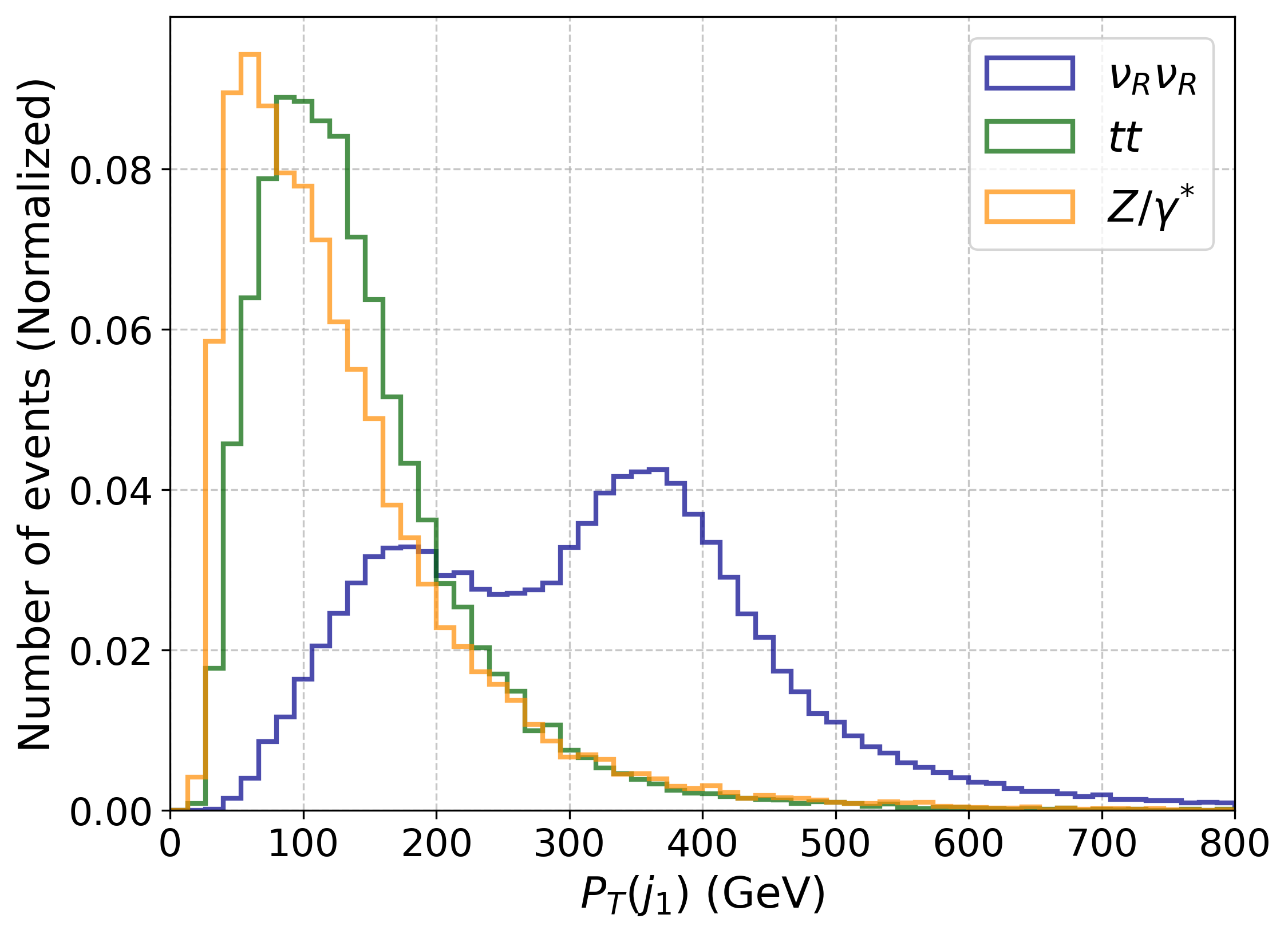}\label{pt_j1_dilep}}\\
    \subfloat[(i)]{\includegraphics[width=0.25\textwidth]{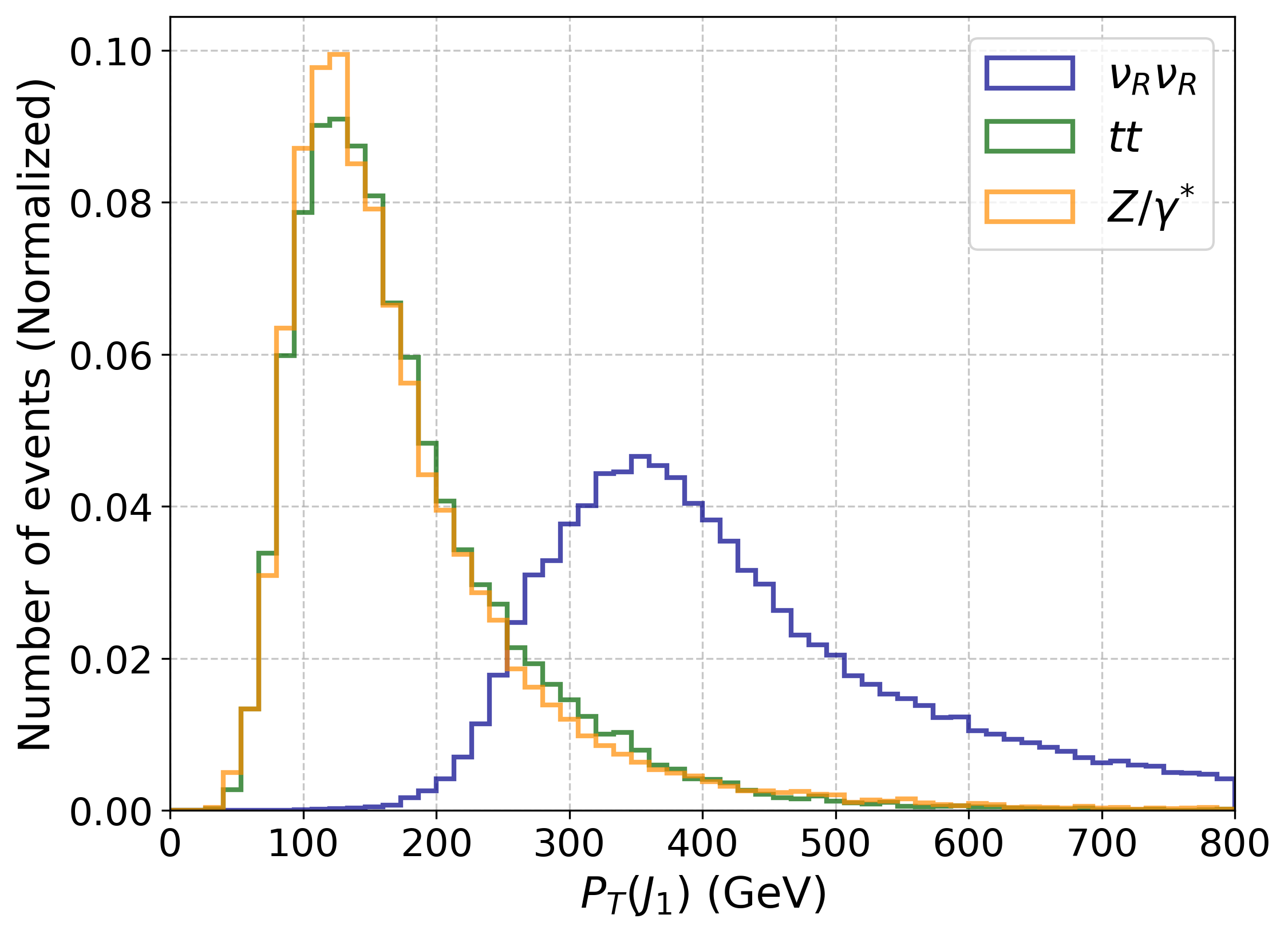}\label{pt_fjt_dilep}}\hfill
    \subfloat[(j)]{\includegraphics[width=0.25\textwidth]{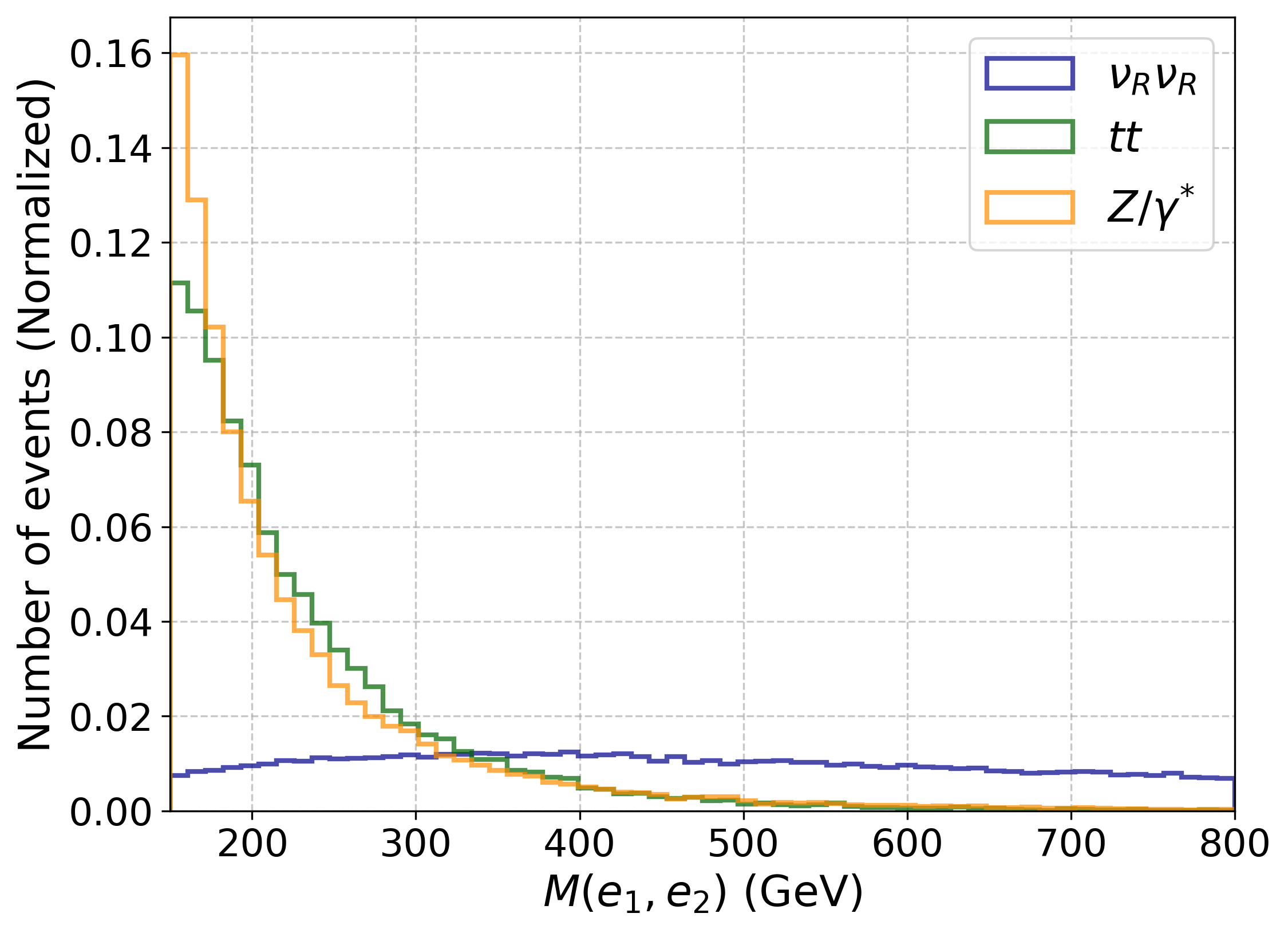}\label{IM_ee_dilep}}\hfill
    \subfloat[(k)]{\includegraphics[width=0.25\textwidth]{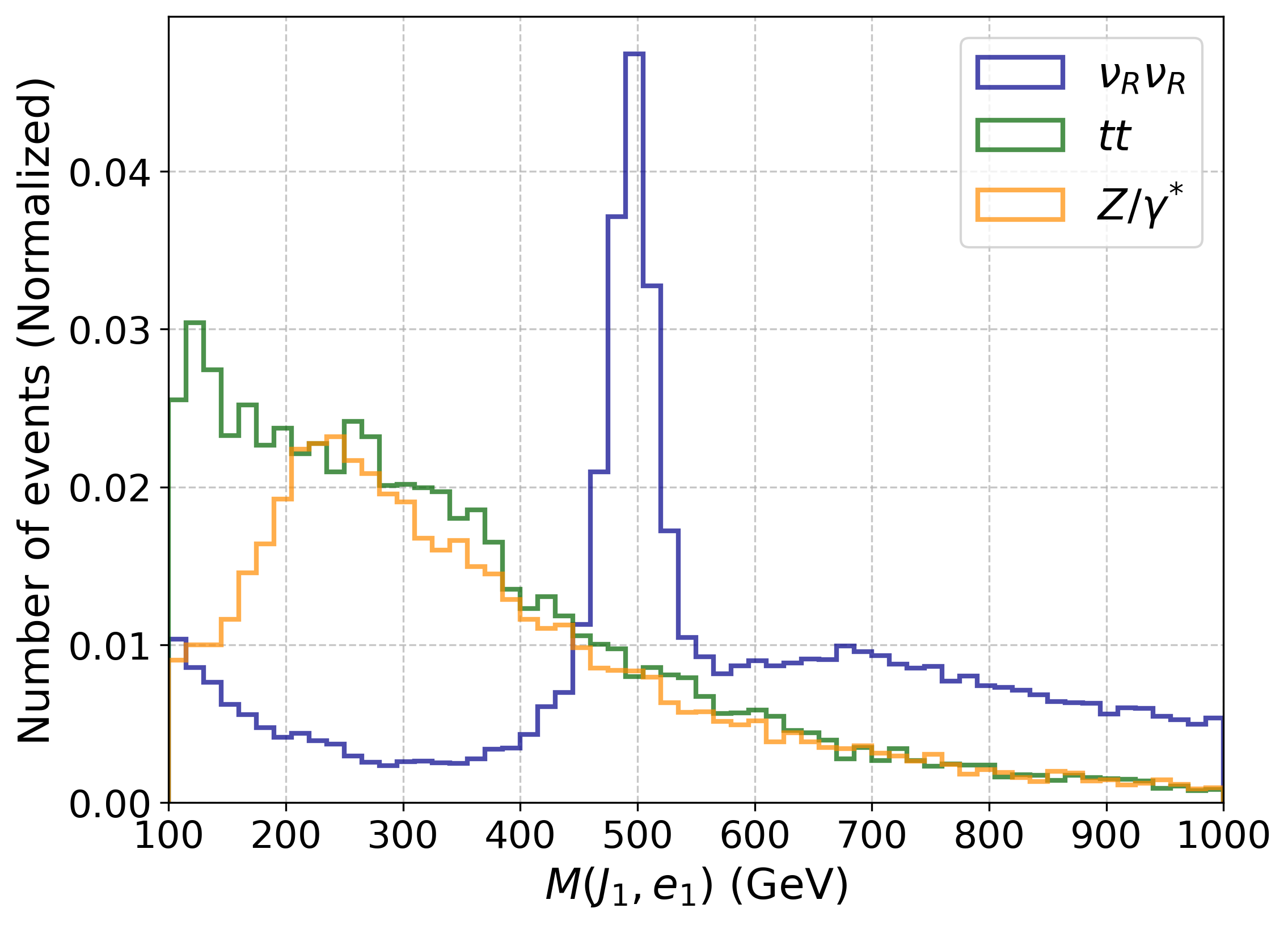}\label{IM_Je_eilep}}\hfill
    \subfloat[(l)]{\includegraphics[width=0.25\textwidth]{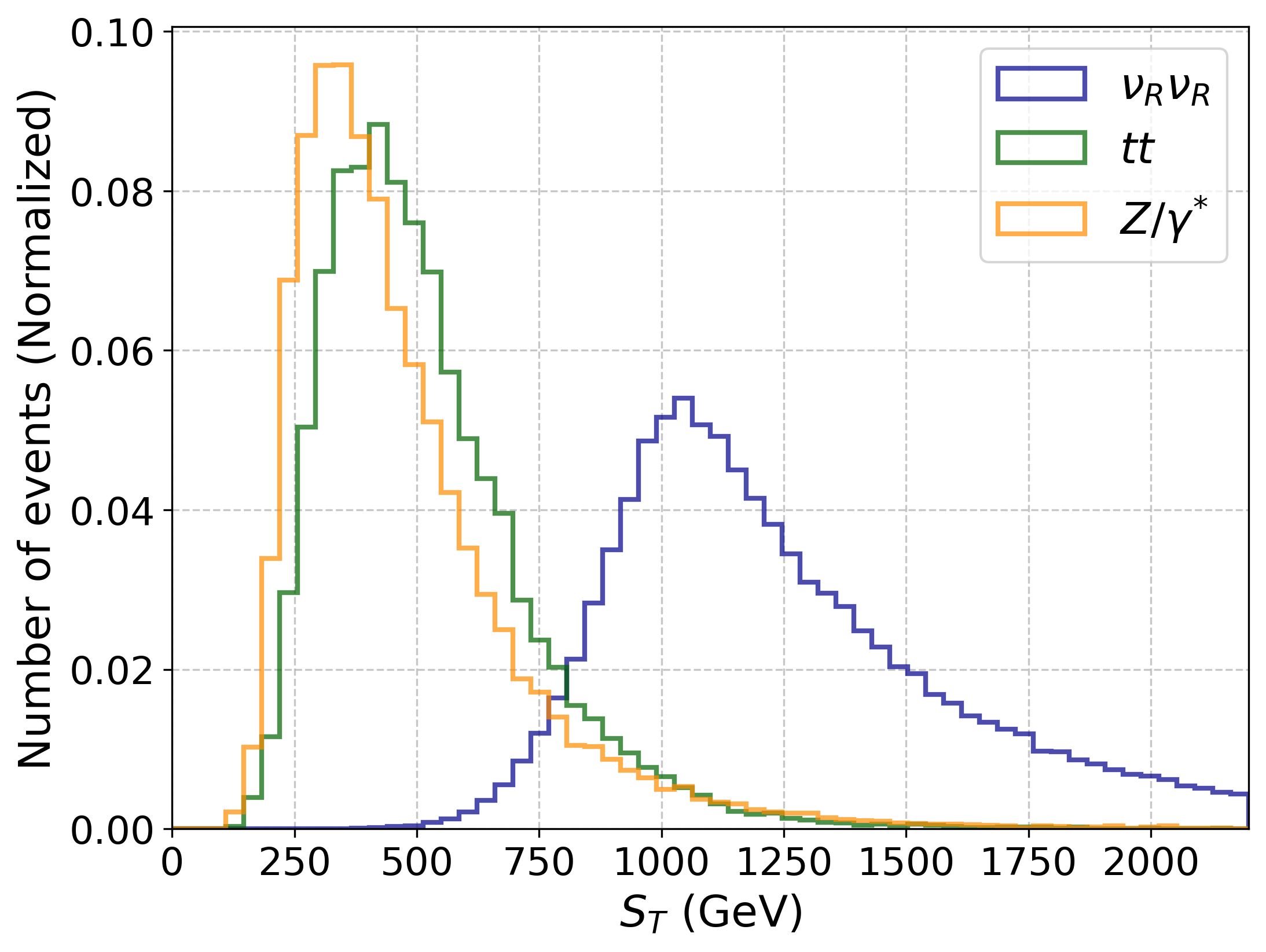}\label{ST_dilep}}
\caption{Normalised distributions of kinematic variables for the signal and two dominant backgrounds in monoelectron [(a)--(f)] and dielectron [(g)--(l)] final states. The signal benchmark is ($M_{U_{1}}, M_{\nu_{R}}$) = (2.5~TeV, 0.5~TeV). For monoelectron, the dominant backgrounds are $W_{e}$ and $t_{e}t_{h}$ and  for dielectron, they are $(Z/\gamma^{*})_e$ and $t_{e}t_{e}$. This distributions are obtained after applying the following cuts at the parton level: for monoelectron $p_{T} (e) > 200$ GeV and for dielectron $M(e_1,e_2) > 150$ GeV.}
    \label{dist}
\end{figure*}

\subsection{Production at the LHC}
\noindent
LQs can be pair-produced via QCD or QED interactions alone, a mixture of QCD-QED interactions~\cite{Bhaskar:2023ftn}, or the RHN-LQ-quark interactions (parametrised by the couplings, $x$ and $y$). Similarly, a single LQ can be produced in association with a RHN through $x$ or $y$. In IP, a LQ is exchanged in the $t$-channel, producing two RHNs from two quarks (see Fig.~\ref{fig:Feyn}). In the high-mass region, this nonresonant $t$-channel production of LQs surpasses their resonant productions when the new coupling(s) involved is(are) sufficiently large. While the PP is largely insensitive to the new couplings (as long as they are not too large), the SP and IP cross sections scale as their second and fourth powers, respectively (see Fig.~\ref{MLQvsCS}). Because of relatively larger PDFs of the first-generation quarks, the IP is more significant for first-generation LQs than the second-generation ones~\cite{Bhaskar:2023xkm}.

In this analysis, we focus on first-generation LQs with $M_{\ell q} > M_{\nu_R}$, allowing them to decay into the first-generation RHN and a jet. As mentioned earlier, we assume this decay to be exclusive, i.e., the BR($\ell_q \to \nu_R q$)$=100$\%. All three production modes of LQs can produce a pair of RHNs accompanied by jets (from LQ decays or radiations). We further assume that the RHN mass is about a few hundred GeV, allowing it to decay resonantly through the $W+e$, $Z+\nu_e$, and $H+\nu_e$ channels (with BRs approximately in $2:1:1$ proportion). Based on how the $\nu_R$ pair decays, we categorise the final states as follows.
\medskip 

\begin{table*}[]
\caption{Cutflow for various background and signal contributions in the monoelectron and dielectron final states. We present the number of events surviving the cuts in Table~\ref{tab:Cuts} for the HL-LHC ($14$~TeV, $3$~ab$^{-1}$; in parentheses, we show the corresponding numbers for vLQs). Two benchmark points are considered to highlight the dominance of LQ IP at higher LQ masses. \label{tab:nof}}
\centering
\renewcommand\baselinestretch{1.6}\selectfont
\begin{tabular*}{\textwidth}{@{\extracolsep{\fill}} lcccc}
\hline
\multirow{2}{*}{\textbf{Monoelectron final state}} & \multicolumn{4}{c}{Selection Cuts} \\ \cline{2-5} 
                                        & C1         & C2                   & C3       & C4       \\ \hline \hline
Signal benchmarks                       &            &           &       &          \\ \hline
$M_{S_1 (U_1)}=1500$ GeV, $M_{\nu_R}= 500$ GeV (LQ PP)                          &   \multicolumn{1}{r}{$30$ ($160$)}         &      \multicolumn{1}{r}{$28$ ($148$)}                &   \multicolumn{1}{r}{$28$ ($148$)}       &   \multicolumn{1}{r}{$22$ ($119$)}       \\
$M_{S_1 (U_1)}=1500$ GeV, $M_{\nu_R}= 500$ GeV (LQ SP)                          &   \multicolumn{1}{r}{$511$ ($4613$)}         &       \multicolumn{1}{r}{$471$ ($4230$)}               &    \multicolumn{1}{r}{$471$ ($4230$)}      &   \multicolumn{1}{r}{$334$ ($3244$)}       \\
$M_{S_1 (U_1)}=1500$ GeV, $M_{\nu_R}= 500$ GeV (LQ IP)                          &   \multicolumn{1}{r}{$184$ ($3209$)}         &        \multicolumn{1}{r}{$133$ ($2292$)}              &    \multicolumn{1}{r}{$133$ ($2292$)}      &    \multicolumn{1}{r}{$82$ ($1437$)}      \\ \hline
                                        &             \multicolumn{3}{l}{Total number of signal events:}       &   \multicolumn{1}{r}{$438$ ($4800$)}       \\ \hline
$M_{S_1 (U_1)}=2500$ GeV, $M_{\nu_R}= 500$ GeV (LQ PP)                          &   \multicolumn{1}{r}{$<1$ ($<1$)}         &       \multicolumn{1}{r}{$<1$ ($<1$)}               &    \multicolumn{1}{r}{$<1$ ($<1$)}      &   \multicolumn{1}{r}{$<1$ ($<1$)}       \\
$M_{S_1 (U_1)}=2500$ GeV, $M_{\nu_R}= 500$ GeV (LQ SP)                          &   \multicolumn{1}{r}{$17$ ($116$)}         &        \multicolumn{1}{r}{$16$ ($111$)}              &    \multicolumn{1}{r}{$16$ ($111$)}      &    \multicolumn{1}{r}{$12$ ($89$)}      \\
$M_{S_1 (U_1)}=2500$ GeV, $M_{\nu_R}= 500$ GeV (LQ IP)                          &    \multicolumn{1}{r}{$50$ ($561$)}        &      \multicolumn{1}{r}{$36$ ($416$)}                &     \multicolumn{1}{r}{$36$ ($416$)}     &   \multicolumn{1}{r}{$24$ ($285$)}       \\ \hline
                                        &            \multicolumn{3}{l}{ Total number of signal events:}        &   \multicolumn{1}{r}{$36$ ($374$)}       \\ \hline
Background processes                      &            &                      &          &          \\ \hline
$W_{e} (+2 j)$                                       & \multicolumn{1}{r}{$1.744 \times 10^6$}  &  \multicolumn{1}{r}{$1.01 \times 10^6$}  & \multicolumn{1}{r}{$1.01 \times 10^6$}  & \multicolumn{1}{r}{$175326$}  \\
$t_{e}t_{h} (+2 j)$                                      & \multicolumn{1}{r}{$387878$}     &   \multicolumn{1}{r}{$302625$}            &   \multicolumn{1}{r}{$30262$}   &    \multicolumn{1}{r}{$50557$}   \\
$W_{e}W_{j} (+2 j)$                                      & \multicolumn{1}{r}{$30830$}      &    \multicolumn{1}{r}{$22383$}            &   \multicolumn{1}{r}{$22383$}   &    \multicolumn{1}{r}{$5791$}      \\
$W_{e}Z_{h} (+2 j)$                                      & \multicolumn{1}{r}{$14474$}      &    \multicolumn{1}{r}{$10889$}            &   \multicolumn{1}{r}{$10889$}   &    \multicolumn{1}{r}{$2382$}      \\
$t_{e}W_{h}+t_{h}W_{e}$                                      &  \multicolumn{1}{r}{$38503$}     &  \multicolumn{1}{r}{$22787$}              &  \multicolumn{1}{r}{$22787$}   &      \multicolumn{1}{r}{$2258$}    \\
$t_{e}b$                                      &   \multicolumn{1}{r}{$6849$}    &        \multicolumn{1}{r}{$5502$}              &      \multicolumn{1}{r}{$5502$}    &  \multicolumn{1}{r}{$831$}        \\
$t_{e}j$                                      &    \multicolumn{1}{r}{$161$}        &         \multicolumn{1}{r}{$131$}             &      \multicolumn{1}{r}{$131$}    &     \multicolumn{1}{r}{$21$}     \\ \hline
                                        &           \multicolumn{3}{l}{ Total number of background events:}     &   \multicolumn{1}{r}{$237166$}       \\ \hline
\multirow{2}{*}{\textbf{Dielectron final state}}   & \multicolumn{4}{c}{Selection Cuts} \\ \cline{2-5} 
                                        & C1         & C2                   & C3       & C4       \\ \hline \hline
Signal benchmarks                       &            &                      &          &          \\ \hline
$M_{S_1 (U_1)}=1500$ GeV, $M_{\nu_R}= 500$ GeV (LQ PP)                          &    \multicolumn{1}{r}{$15$ ($136$)}        &    \multicolumn{1}{r}{$15$ ($136$)}                  &       \multicolumn{1}{r}{$13$ ($125$)}   &   \multicolumn{1}{r}{$13$ ($125$)}       \\
$M_{S_1 (U_1)}=1500$ GeV, $M_{\nu_R}= 500$ GeV (LQ SP)                          &    \multicolumn{1}{r}{$285$ ($2768$)}        &           \multicolumn{1}{r}{$284$ ($2764$)}           &   \multicolumn{1}{r}{$230$ ($2522$)}       &   \multicolumn{1}{r}{$228$ ($2516$)}       \\
$M_{S_1 (U_1)}=1500$ GeV, $M_{\nu_R}= 500$ GeV (LQ IP)                          &    \multicolumn{1}{r}{$182$ ($2478$)}        &           \multicolumn{1}{r}{$177$ ($2317$)}           &   \multicolumn{1}{r}{$150$ ($2016$)}       &    \multicolumn{1}{r}{$117$ ($1518$)}      \\ \hline
                                        &           \multicolumn{3}{l}{ Total number of signal events: }        &   \multicolumn{1}{r}{$358$ ($4159$)}       \\ \hline
$M_{S_1 (U_1)}=2500$ GeV, $M_{\nu_R}= 500$ GeV (LQ PP)                          &      \multicolumn{1}{r}{$<1$ ($<1$)}      &              \multicolumn{1}{r}{$<1$ 
($<1$)}      &   \multicolumn{1}{r}{$<1$ ($<1$)}       &   \multicolumn{1}{r}{$<1$ ($<1$)}       \\
$M_{S_1 (U_1)}=2500$ GeV, $M_{\nu_R}= 500$ GeV (LQ SP)                          &     \multicolumn{1}{r}{$11$ ($91$)}       &      \multicolumn{1}{r}{$11$ ($91$)}                &    \multicolumn{1}{r}{$9$ ($86$)}      &    \multicolumn{1}{r}{$9$ ($86$)}      \\
$M_{S_1 (U_1)}=2500$ GeV, $M_{\nu_R}= 500$ GeV (LQ IP)                          &      \multicolumn{1}{r}{$40$ ($441$)}      &      \multicolumn{1}{r}{$38$ ($423$)}                &    \multicolumn{1}{r}{$33$ ($372$)}      &   \multicolumn{1}{r}{$26$ ($310$)}       \\ \hline
                                        &           \multicolumn{3}{l}{ Total number of signal events:}        &  \multicolumn{1}{r}{$35$ ($396$)}        \\ \hline
Background processes                      &            &                      &          &          \\ \hline
$(Z/\gamma^{*})_{e} (+2 j)$                                       &  \multicolumn{1}{r}{$253822$}          &    \multicolumn{1}{r}{$98786$}                  &   \multicolumn{1}{r}{$51136$}       &  \multicolumn{1}{r}{$9065$}        \\
$t_{e}t_{e} (+2 j)$                                      &     \multicolumn{1}{r}{$23964$}       &    \multicolumn{1}{r}{$10523$}                  &      \multicolumn{1}{r}{$6301$}    &   \multicolumn{1}{r}{$622$}       \\
$W_{e}W_{e} (+2 j)$                                      &      \multicolumn{1}{r}{$6159$}      &      \multicolumn{1}{r}{$3216$}                &      \multicolumn{1}{r}{$1909$}    &    \multicolumn{1}{r}{$494$}      \\
$W_{h}Z_{e} (+2 j)$                                      &   \multicolumn{1}{r}{$772$}         &     \multicolumn{1}{r}{$519$}                 &      \multicolumn{1}{r}{$324$}    &    \multicolumn{1}{r}{$138$}      \\
$t_{e}W_{e}$                                     &      \multicolumn{1}{r}{$2276$}      &    \multicolumn{1}{r}{$981$}                  &     \multicolumn{1}{r}{$470$}     &    \multicolumn{1}{r}{$83$}      \\
$Z_{e}Z_{h} (+2 j)$                                      &    \multicolumn{1}{r}{$130$}        &       \multicolumn{1}{r}{$63$}               &      \multicolumn{1}{r}{$32$}    &    \multicolumn{1}{r}{$8$}      \\ \hline
                                        &            \multicolumn{3}{l}{ Total number of background events: }     &   \multicolumn{1}{r}{$10410$}       \\ \hline
\end{tabular*}
\end{table*}
\begin{table*}
\centering
\caption{Model-independent (i.e., from QCD-driven PP contributions obtained by taking $x\to 0$) mass limits on vLQs in GeVs for two different $\kappa$ values. These numbers are obtained for the HL-LHC.\label{tab:kappa}}
\small
\renewcommand\baselinestretch{1.5}\selectfont
\begin{tabular*}{\textwidth}{@{\extracolsep{\fill}} lllllllll}
\hline
\multirow{3}{*}{vLQ} & \multicolumn{4}{c}{Monoelectron}                         & \multicolumn{4}{c}{Dielectron}                          \\ \cline{2-9} 
                    & \multicolumn{2}{c}{$\kappa= 1$} & \multicolumn{2}{c}{$\kappa= 0$} & \multicolumn{2}{c}{$\kappa= 1$} & \multicolumn{2}{c}{$\kappa= 0$} \\ \cline{2-9} 
                    & $5\sigma$       & $2\sigma$      & $5\sigma$       & $2\sigma$      & $5\sigma$      & $2\sigma$      & $5\sigma$      & $2\sigma$      \\ \hline \hline
$U_1$  &       $1160$      &      $1190$        &    $1315$         &      $1450$  &   $1269$     &  $1373$     &   $1553$   &  $1697$                     \\ \hline
$\overline{U}_1$               &      $1160$         &  $1190$           &   $1315$           &  $1450$           &          $1269$    &      $1373$       &      $1553$       &      $1697$        \\ \hline
$\widetilde{V}_2$                  &     $1185$          &       $1300$      &        $1400$      &      $1622$       &         $1346$     &         $1537$    &       $1651$      &       $1833$       \\ \hline
\end{tabular*}
\end{table*}

\noindent
\textbf{Monoelectron}: For the final state to have only one electron and no other charged lepton, one $\nu_R$ has to decay via the $W+e$ channel and the other one to $Z+\nu_e$ or $H+\nu_e$ channels where the $Z/H$ boson decays hadronically (which can produce a fatjet). The various production channels contributing to monoelectron final states are as follows (see Fig.~\ref{fig:Feyn}):
\begin{align*}
    pp\ \to\ \left\{\begin{array}{l}
    \ell_q\ell_q\\
    \ell_q\,\nu_{R}\ (+j)\\
    \nu_R\nu_{R}\ (+j)
    \end{array}\right\}\to& \left\{\begin{array}{l}
    (j\nu_{R})(j\nu_{R})\\
    (j\nu_{R})\nu_{R}\ (+j)\\
    \nu_R\nu_{R}\ (+j)\\
    \end{array}\right\}\nn\\\to&~\lt(e^{\pm} W_{h}^{\mp}\rt) 
    \lt(\n_e Z_{h}/H_h\rt)+ \mbox{jet(s)},
\end{align*}
with the $\n_e$ leading to missing transverse energy (MET). The field subscripts (except for $\ell_q$ and $\n_{L/R}$) indicate decay modes; for example, $h$ implies hadronic decay. The most dominant channel has a pair of RHN decaying as $\nu_R \nu_R \to (e^{\pm} W_h^{\mp}) (\nu_{e} Z_h/H_h)$, with an overall BR of about $23$\%. The second most dominant channel, \(\nu_R \nu_R \to (e^{\pm} W_h^{\mp}) (\nu_e Z_{\nu_\ell})\), has a BR of about $7$\%. However, its contribution to our final results is minimal due to the low selection efficiency. All other channels with \( \mathrm{BR} \lesssim 1\% \) have a negligible impact on the results.
\medskip

\noindent
\textbf{Dielectron}: An electron-positron pair ($e^+e^-$) is produced when both RHNs decay via the $W+e$ channel. The final state with an $e^-e^+$ pair and no other charged lepton does not involve MET, making it fully reconstructible. The hadronic decays of the two $W$ bosons can result in two fatjets. The various production channels contributing to dielectron final states are as follows:
\begin{align*}
    pp\ \to\ \left\{\begin{array}{l}
    \ell_q\ell_q\\
    \ell_q\,\nu_{R}\ (+j)\\
    \nu_R\nu_{R}\ (+j)
    \end{array}\right\}&\to \left\{\begin{array}{l}
    (j\nu_{R})(j\nu_{R})\\
    (j\nu_{R})\nu_{R}\ (+j)\\
    \nu_R\nu_{R}\ (+j)\\
    \end{array}\right\}\nn\\&\to~ e^{\pm} W_{h}^{\mp} e^{\mp} W_{h}^{\pm}+ \mbox{jet(s)}.
\end{align*}
In the dielectron channel, the dominant decay chain is $\nu_R \nu_R \to (e^{\pm} W_h) (e^{\mp} W_h)$, with an overall BR of $22$\%. Other channels, such as $\nu_R \nu_R \to (\nu_{e} Z_h/H_h) (\nu_{e} Z_{e})$, have BRs of the order of $1$\%, and contribute negligibly to the results. A $Z$-veto to suppress the Drell-Yan dilepton background effectively eliminates the contribution from these subdominant processes. Consequently, we focus on the most dominant signals. 

There are existing searches for dilepton final states via RHNs~\cite{Chakdar95,Ng:2015hba,Das:2017flq,Das:2017deo,Cox:2017eme}. These studies are primarily motivated by Type-I seesaw models, where the RHNs are fully Majorana-type fermions leading to same-sign dilepton final states with lepton number violation. In contrast, we rely on the framework of inverse seesaw mechanism, where RHNs are pseudo-Dirac in nature, resulting in opposite-sign dilepton final states.
\medskip

\noindent
\textbf{Trilepton}: Trilepton final states through RHN decays can arise via the decay of a heavy neutral gauge boson or a scalar.  Both symmetric and asymmetric decay of the RHN pair can lead to the trilepton signal. The production and decay processes leading to these final states are as follows:
\begin{align*}
    pp\ \to\ \left\{\begin{array}{l}
    \ell_q\ell_q\\
    \ell_q\,\nu_{R}\ (+j)\\
    \nu_R\nu_{R}\ (+j)
    \end{array}\right\}\to& \left\{\begin{array}{l}
    (j\nu_{R})(j\nu_{R})\\
    (j\nu_{R})\nu_{R}\ (+j)\\
    \nu_R\nu_{R}\ (+j)\\
    \end{array}\right\}\nn\\\to&\left\{\begin{array}{l}
    (e^{\pm} W_{\ell}^{\mp}) (e^{\pm} W_{h}^{\pm}) + \mbox{jet(s)}\\
    (e^{\pm} W_{h}^{\mp}) (\n_eZ_{\ell})+ \mbox{jet(s)}\end{array}\right\}.
\end{align*}
Even if we consider the RHN to be first generation, a leptonically decaying vector boson can decay to a $\m$ or a $\ta$ lepton to produce a (mixed-flavour) trilepton final state. Trilepton channels have been explored in detail in Refs.~\cite{Kang16,Accomando:2017qcs,Helo:2018rll}.
\medskip

\noindent
\textbf{Quadlepton}: A pair of RHNs can also decay to give four-lepton final states as:
\begin{align*}
    pp\ \to\ \left\{\begin{array}{l}
    \ell_q\ell_q\\
    \ell_q\,\nu_{R}\ (+j)\\
    \nu_R\nu_{R}\ (+j)
    \end{array}\right\}\to& \left\{\begin{array}{l}
    (j\nu_{R})(j\nu_{R})\\
    (j\nu_{R})\nu_{R}\ (+j)\\
    \nu_R\nu_{R}\ (+j)\\
    \end{array}\right\}\nn\\\to&\left\{\begin{array}{l}
    e^{\pm} W_{\ell}^{\mp} e^{\mp} W_{\ell}^{\pm} + \mbox{jet(s)}\\
    \nu_{e} Z_{\ell} \nu_{e} Z_{\ell}+ \mbox{jet(s)}\end{array}\right\}.
\end{align*}
\noindent
This channel suffers from the small leptonic BRs of the heavy bosons. However, it might still be accessible at the HL-LHC since the four-lepton SM background is small. Similar channels have been analysed in the context of a $U(1)_{B-L}$ model in Ref.~\cite{Huitu08}.
\medskip

\noindent
\textbf{Displaced vertex}: 
The RHN can exhibit a displaced vertex signature if its decay width is sufficiently small. In that case, the search strategy will differ significantly from that for prompt decays. The decay width of the RHN depends on its mass and the angle of mixing with its lighter counterpart. Displaced vertex signatures have been studied in Refs.~\cite{Frank10,Chiang:2019ajm,DAS2019135052} in the context of $Z'\to \nu_R\nu_R$ decay and in Refs.~\cite{Deppisch:2018eth,Liu:2022ugx} in the context of $\phi\to\nu_R\nu_R$ decay. In extreme scenarios, the decay width of the RHN may become so small that it decays outside the detector, leading to missing energy in the event. Such cases require a separate analysis tailored to identifying missing energy signals.
\medskip

\noindent
\textbf{Fatjet with MET}: There are also nonleptonic final states involving one or more fatjets accompanied by some MET as follows.
\begin{align*}
    pp\ \to\ \left\{\begin{array}{l}
    \ell_q\ell_q\\
    \ell_q\,\nu_{R}\ (+j)\\
    \nu_R\nu_{R}\ (+j)
    \end{array}\right\}\to& \left\{\begin{array}{l}
    (j\nu_{R})(j\nu_{R})\\
    (j\nu_{R})\nu_{R}\ (+j)\\
    \nu_R\nu_{R}\ (+j)\\
    \end{array}\right\}\nn\\\to&\ (\nu_e Z_{h}/H_{h}) (\nu_e Z_{h}/H_{h})+ \mbox{jet(s)}.
\end{align*}

\noindent
The background for this signal is expected to be very large. However, the signal can still be separable using advanced jet-substructure variables and machine-learning techniques. Similar final states have been analysed in the context of the inert Higgs doublet model in Ref.~\cite{Bhardwaj:2019mts}.

\subsection{Signal, background and selection criteria}
\noindent
In this paper, we focus on the monoelectron and dielectron final states. We require at least two AK4 jets and a fatjet in the monoelectron channel, and two electrons and at least one fatjet in the dielectron channel. The relevant SM backgrounds and their cross sections are listed in Table~\ref{tab:Backgrounds}. Table~\ref{tab:Cuts} summarises the cuts applied to isolate the signals from the backgrounds. These are decided from the signal and background kinematic distributions, some of which are shown in Fig.~\ref{dist}. We show the cut-flows (the number of signal and background events surviving the cuts) for two benchmark points in the monoelectron and dielectron channels in Table~\ref{tab:nof}. For a fixed RHN mass, increasing the LQ mass enhances the importance of its IP, which contributes more to the total signal events than the combined contributions of the PP and SP. 

The dominant SM background for monoelectron and dielectron final states are $W_e+$ jets and $(Z/\gamma^*)_e+$ jets, respectively, owing to their large cross-sections. Selection cuts try to suppress these contributions significantly while retaining a substantial fraction of the signal events. Consequently, the cuts are optimised to maximise event captures from the $t$-channel LQ exchange process. (The selection cuts used in this study differ slightly from those applied in Ref.~\cite{Bhaskar:2023xkm}, which focused on RHN production via second-generation LQs. This difference arises mainly because the $t$-channel LQ exchange plays a more significant role than the PP and the SP for first-generation LQs than the second-generation ones.)  The next most significant background (for both channels) is $tt+$ jets. As long as the signal contains no $b$ jets, a $b$ veto can tame it. In the monoelectron channel, a significant contribution comes from the $\nu_R \to H\nu_e \to bb\nu_e$ decay. Hence, we impose a $b$ veto only in the dilepton mode.

The fatjets in the signal processes primarily originate from $W/Z$ decays, with the leading fatjet mass peaking around their masses. Consequently, applying a window-like cut around this peak can effectively reduce background contributions. However, the fatjet mass distribution for the signal exhibits a bimodal nature, with a second peak appearing at the RHN mass. Due to this characteristic, we refrain from imposing a cut on the fatjet mass in our analysis.

\begin{figure*}[]
    \centering
    \captionsetup[subfigure]{labelformat=empty}
    \subfloat[(a)]{\includegraphics[width=0.225\textwidth]{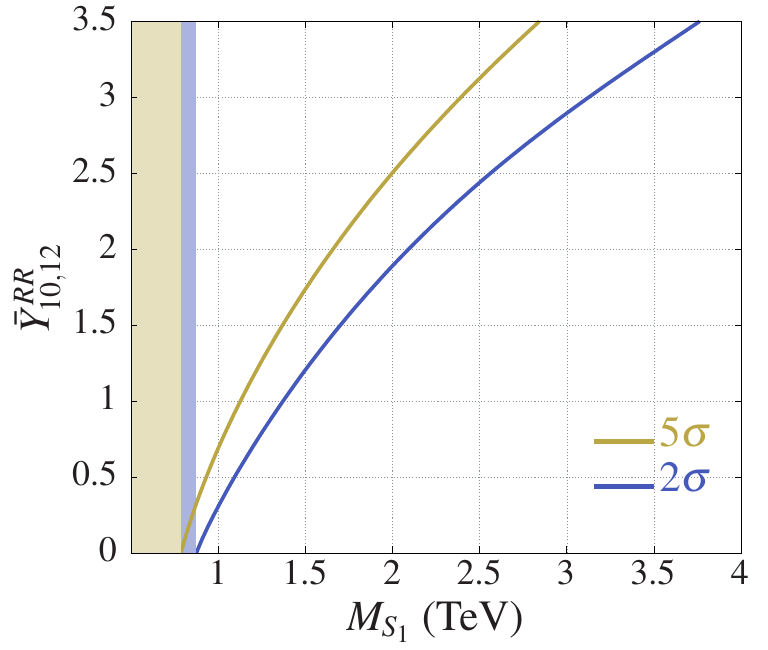}\label{S1_mono}}\hspace{0.5cm}
    \subfloat[(b)]{\includegraphics[width=0.225\textwidth]{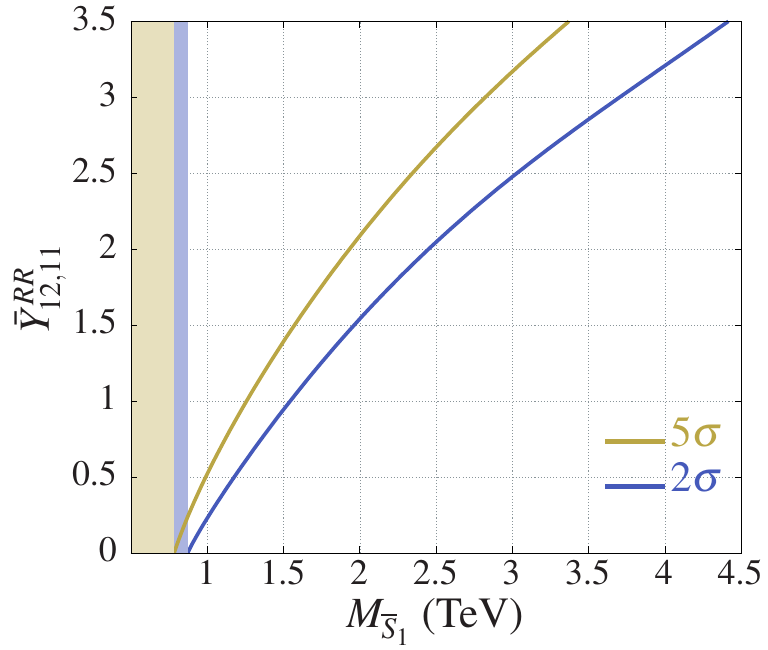}\label{S1bar_mono}}\hspace{0.5cm}
    \subfloat[(c)]{\includegraphics[width=0.225\textwidth]{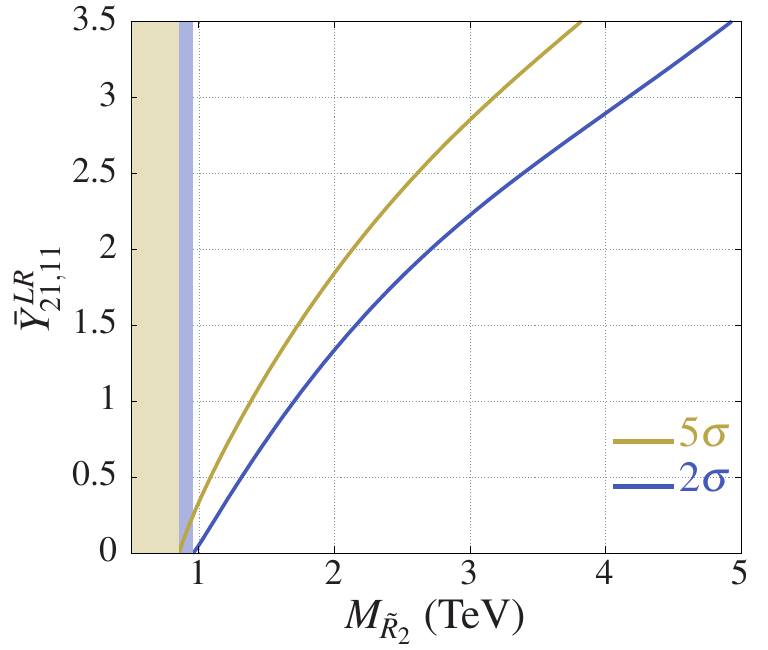}\label{R2t_mono}}\hspace{0.5cm}
    \subfloat[(d)]{\includegraphics[width=0.225\textwidth]{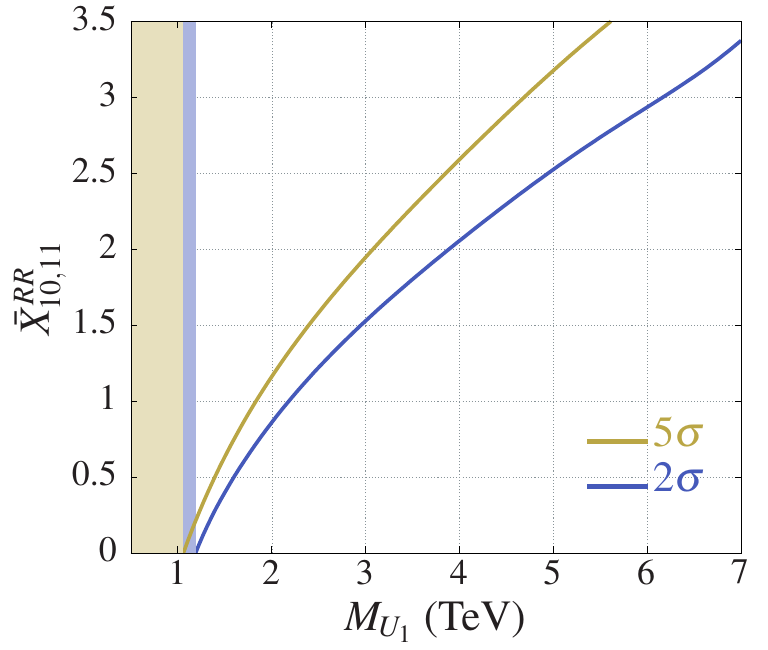}\label{U1k0_mono}}\\
    \subfloat[(e)]{\includegraphics[width=0.225\textwidth]{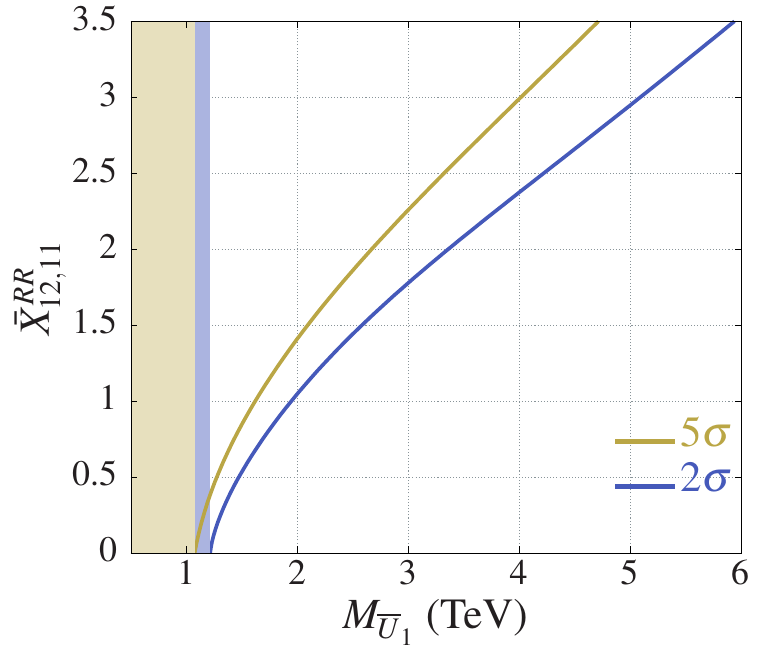}\label{U1bar_mono}}\hspace{0.5cm}
    \subfloat[(f)]{\includegraphics[width=0.225\textwidth]{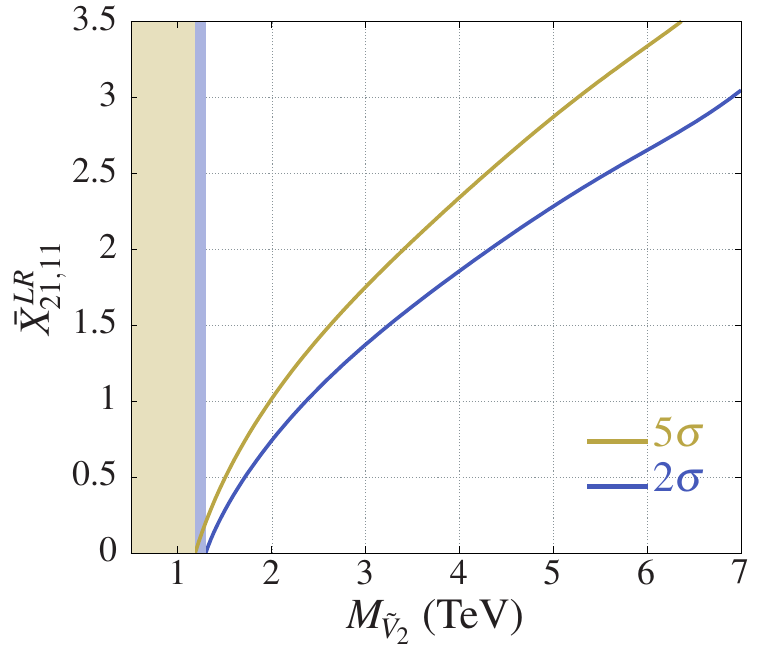}\label{V2t_mono}}\hspace{0.5cm}
    \subfloat[(g)]{\includegraphics[width=0.225\textwidth]{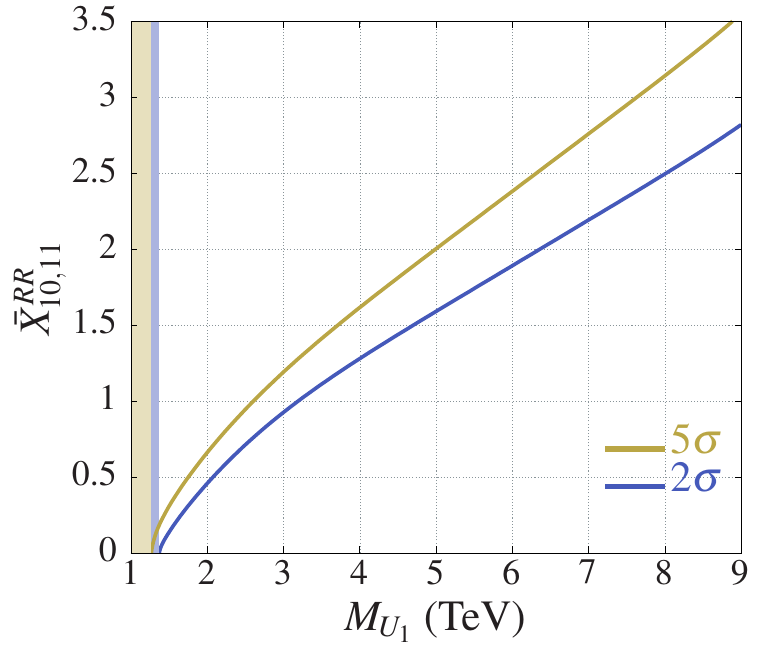}\label{U1_di}}\hspace{0.5cm}
    \subfloat[(h)]{\includegraphics[width=0.225\textwidth]{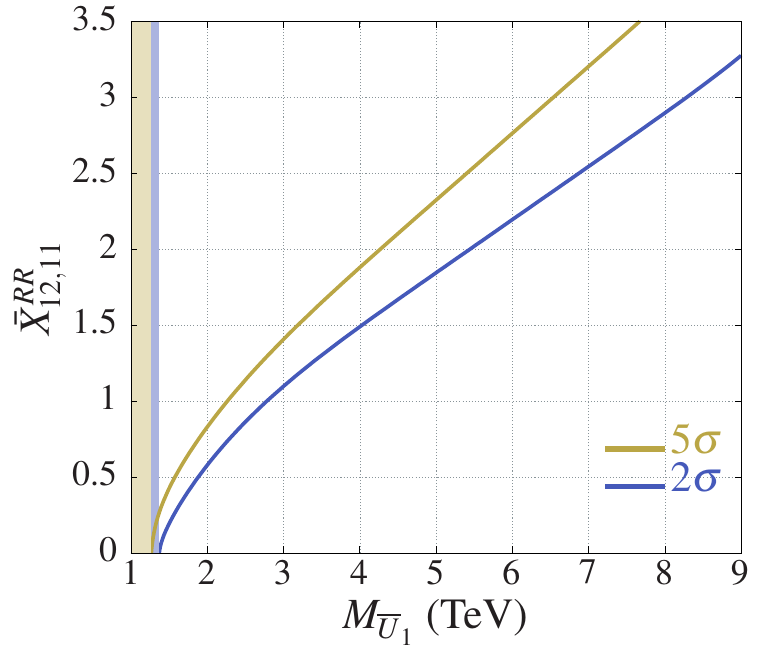}\label{U1bar_di}}\\
    \subfloat[(i)]{\includegraphics[width=0.225\textwidth]{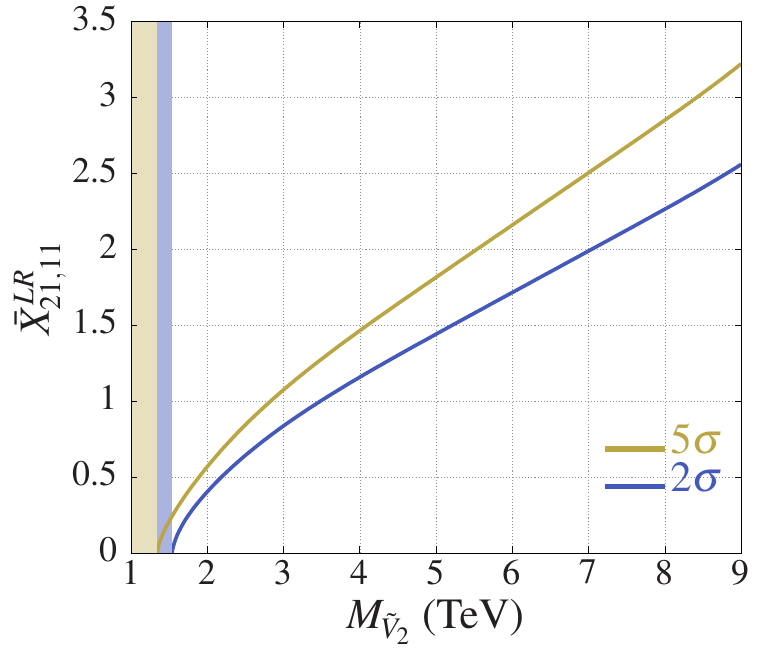}\label{V2t_di}}\hspace{0.5cm}
    \subfloat[(j)]{\includegraphics[width=0.225\textwidth]{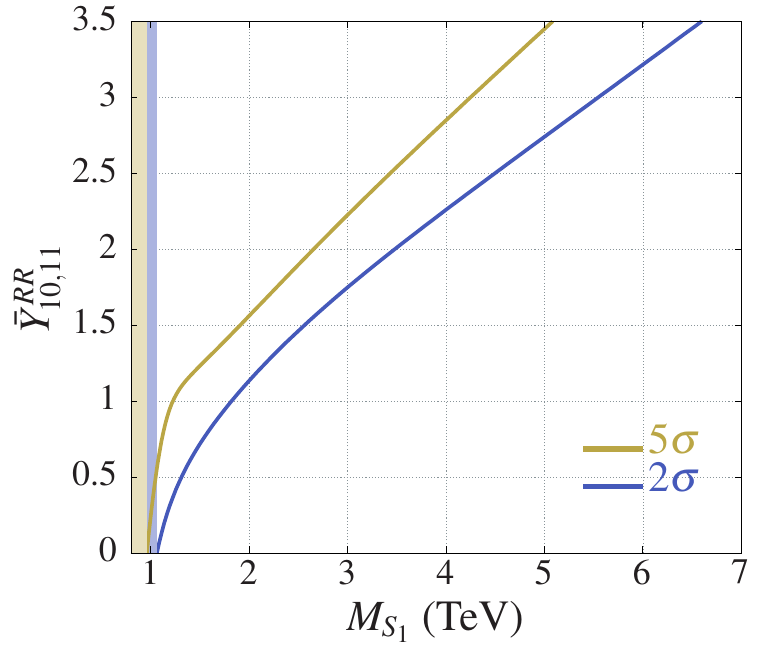}\label{S1_di}}\hspace{0.5cm}
    \subfloat[(k)]{\includegraphics[width=0.225\textwidth]{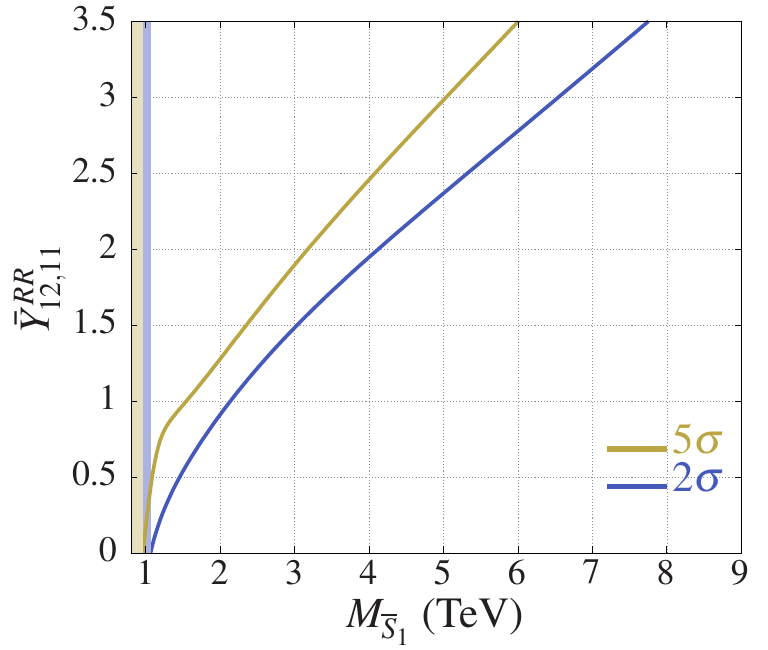}\label{S1bar_di}}\hspace{0.5cm}
    \subfloat[(l)]{\includegraphics[width=0.225\textwidth]{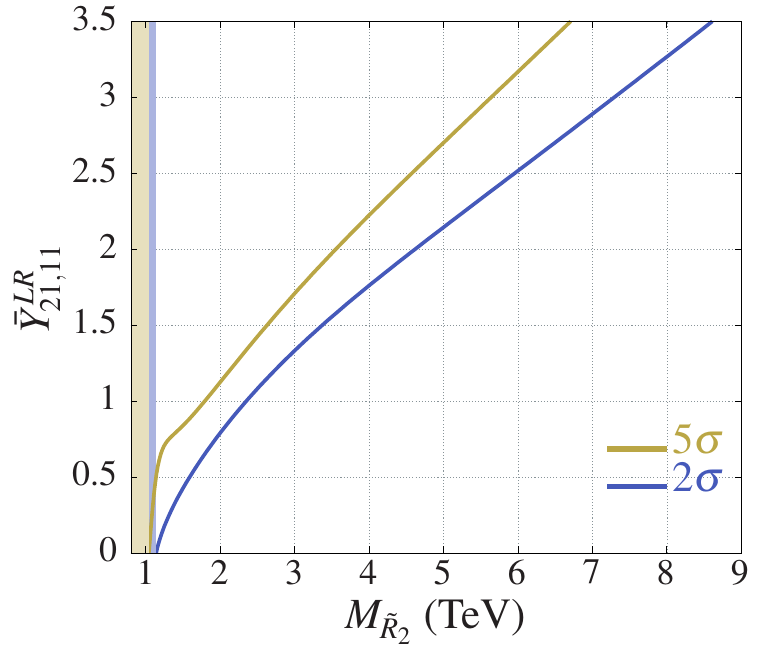}\label{R2t_di}}
    \caption{The $2\sigma$ (exclusion) and $5\sigma$ (discovery) regions for sLQs and vLQs at the HL-LHC: [(a) -- (f)] for monoelectron final states and [(g) -- (l)] for dielectron final states. The vertical shaded regions indicate the model-independent limits (i.e., obtained only with PP by setting $x/y\to 0$) for different LQs. All these limits are obtained for $M_{\nu_R} = 500$ GeV. We set $\kappa=1$ in the vLQ plots.}
    \label{MLQvslam}
    
    \subfloat[(a)]{\includegraphics[width=0.25\textwidth]{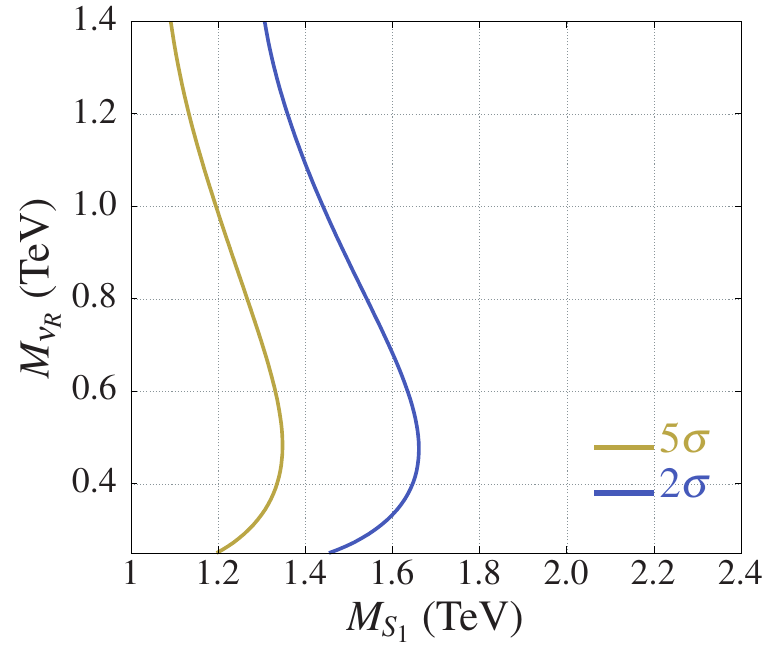}\label{S1_rhn}}\hspace{0.5cm}
    \subfloat[(b)]{\includegraphics[width=0.25\textwidth]{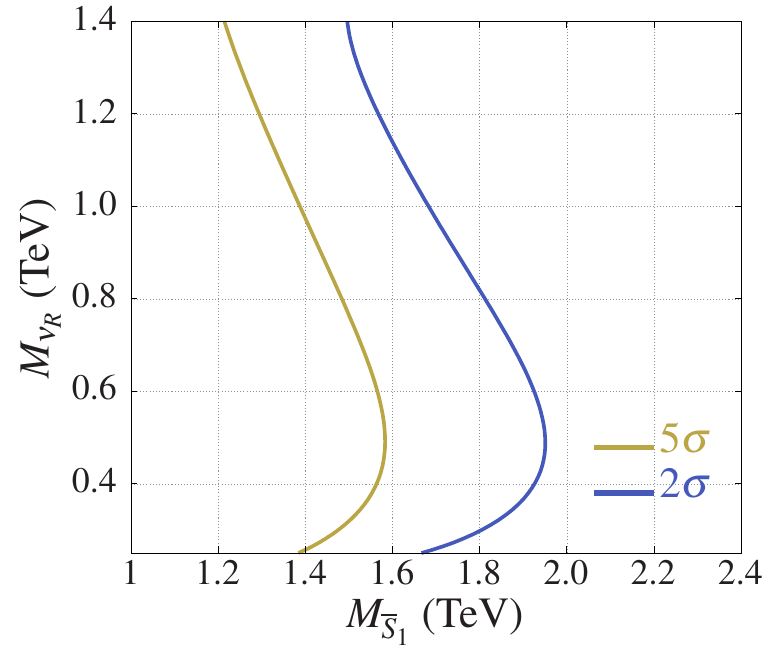}\label{S1bar_rhn}}\hspace{0.5cm}
    \subfloat[(c)]{\includegraphics[width=0.25\textwidth]{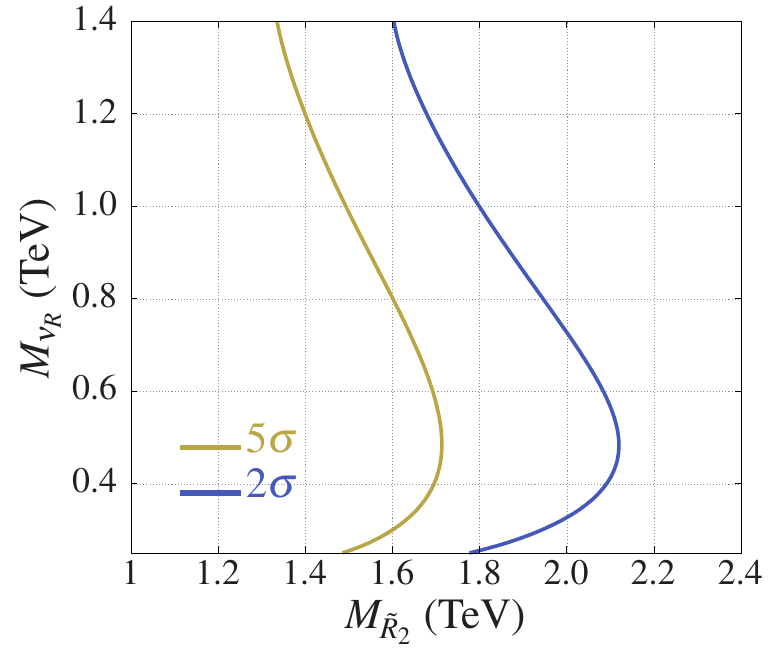}\label{R2t_rhn}}\\
    \subfloat[(d)]{\includegraphics[width=0.25\textwidth]{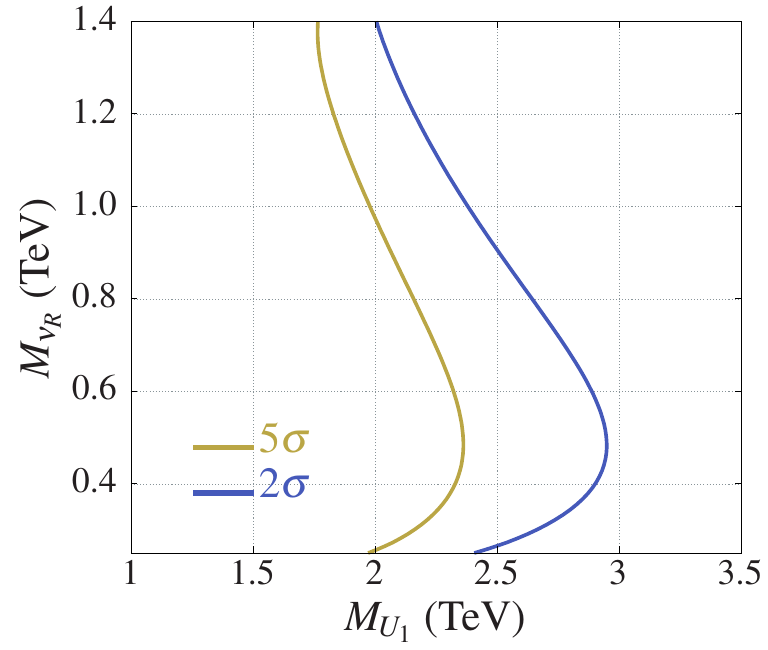}\label{U1_rhn}}\hspace{0.5cm}
    \subfloat[(e)]{\includegraphics[width=0.25\textwidth]{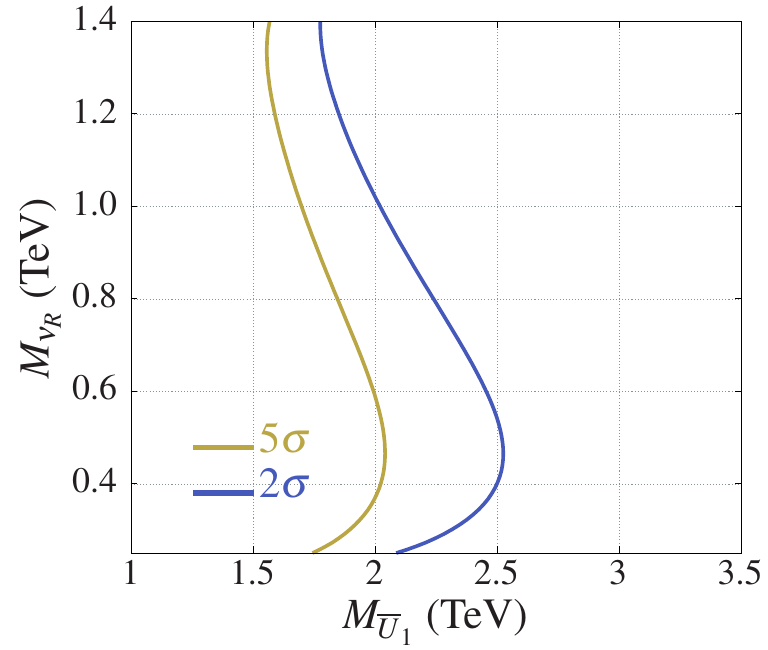}\label{U1bar_rhn}}\hspace{0.5cm}
    \subfloat[(f)]{\includegraphics[width=0.25\textwidth]{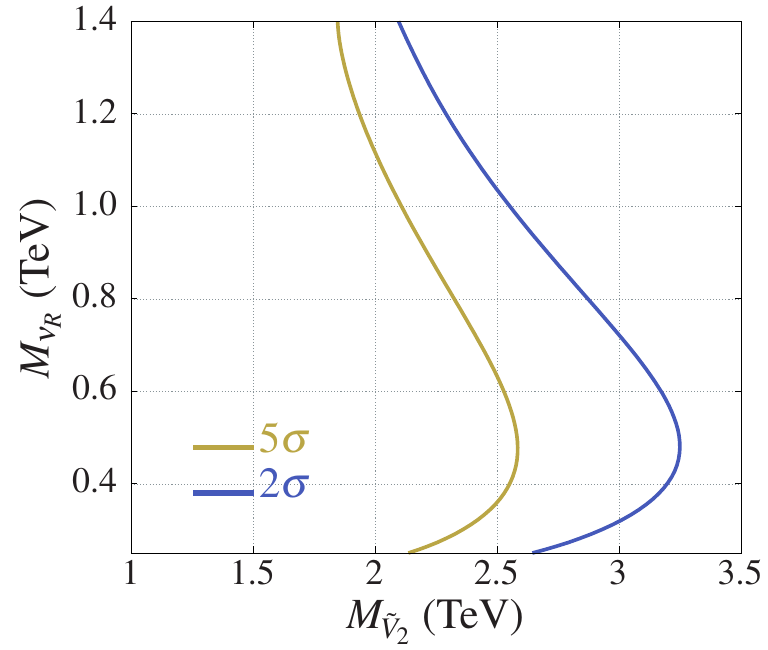}\label{V2t_rhn}}    
    \caption{The $2\sigma$ and $5\sigma$ regions for the dielectron final states on the $M_{\ell_q}-M_{\nu_R}$ planes with fixed coupling, $x/y = 1$.}\label{LQvsRHN}
\end{figure*}

\section{HL-LHC prospects} \label{sec:HLLHCpros}
\noindent
For the HL-LHC ($14$~TeV centre-of-mass energy with $\mathfrak{L}=3$~ab$^{-1}$ of integrated luminosity), we estimate the number of signal events ($N_S$) and number of total background events ($N_B$) surviving the cuts listed in Table~\ref{tab:Cuts}. The $N_S$ is calculated as,
\begin{equation}
N_{\text{S}} = \Big( \sigma_{\text{PP}} \cdot \epsilon_{\text{PP}} + \lambda^2 \cdot \sigma_{\text{SP}} \cdot \epsilon_{\text{SP}} + \lambda^4 \cdot \sigma_{\text{IP}} \cdot \epsilon_{\text{IP}} \Big) \cdot \mathfrak{L},
\end{equation}
where \(\sigma_{XX}\) and \(\epsilon_{XX}\) ($XX=\text{PP},~\text{SP},~\text{IP}$) are the cross sections and efficiencies of the different signal processes. Here, \(\lambda\) represents the LQ coupling (\(x\) for sLQs and \(y\) for vLQs). Both \(\sigma\) and \(\epsilon\) are functions of \(M_{\ell_q}\) and \(M_{\nu_R}\), making \(N_{\text{S}}\) a function of these masses and \(\lambda\). To determine the signal sensitivity, we use the following \(\mathcal{Z}\)-score~\cite{Cowan:2010js} formula:  
\begin{equation}
\label{eq:Zscore}
\mathcal{Z} = \sqrt{2\left(N_{\text{S}} + N_{\text{B}}\right)\ln\left(\frac{N_{\text{S}} + N_{\text{B}}}{N_{\text{B}}}\right) - 2N_{\text{S}}}\,.
\end{equation}
 
The results for the monoelectron and dielectron final states are presented in Fig.~\ref{MLQvslam}, showing $2\sigma$ ($\sim$ exclusion) and $5\sigma$ (discovery) contours in the $M_{\ell q}-x(y)$ plane for sLQs (vLQs). For these plots, we set the benchmark RHN mass at $500$~GeV. The dielectron channel offers stricter exclusions than the monoelectron channel because a cut on the dielectron invariant mass can significantly suppress the dominant Drell-Yan background. In Fig.~\ref{LQvsRHN}, we show similar contours on the $M_{\ell_q}-M_{\nu_R}$ plane for $\lambda=1$ in the dielectron channel. With increasing \( M_{\nu_R} \), the contours first bulge towards higher $M_{\ell_q}$ values till about $M_{\nu_R}=500$~GeV and then return towards lower $M_{\ell_q}$ values. This happens because the selection efficiencies start low at low $M_{\nu_R}$ values and increase till about $M_{\nu_R}=500$~GeV, where the cuts are optimised. Beyond this point, the efficiencies plateau, causing the contours to move towards smaller \( M_{\ell_q} \) values as expected. The exclusion limits for vLQs are largely insensitive to $\kappa$, as the dominant contribution arises from the IP process, which is independent of $\kappa$. Although PP and SP depend on $\kappa$, their overall impact is relatively smaller. Consequently, for larger LQ masses where the IP process dominates, the effect of $\kappa$ on limits remains negligible. The model-independent mass limits (from PP with $x\to 0$) for different vLQs for $\kappa=0,1$ are presented in Table~\ref{tab:kappa}. (Note that since we optimise the cuts for the IP process, we obtain slightly less stringent model-independent mass limits than those obtained in Ref.~\cite{Bhaskar:2023xkm}.) 

\section{Summary and Conclusions}
\label{sec:conclu}

\noindent
We investigated RHN productions at the LHC mediated by first-generation LQs and demonstrated the importance of the $t$-channel LQ exchange process in this case. The first-generation process shows more promise at the HL-LHC than the second-generation one~\cite{Bhaskar:2023xkm} because the larger quark PDFs significantly boost this process.  We designed a set of criteria/cuts optimised for this channel. Though the cuts are less stringent than those used earlier in the second-generation case~\cite{Bhaskar:2023xkm}, the enhanced cross-sections result in stricter exclusion limits. 

We estimated the projected HL-LHC exclusion limits and discovery reaches for all first-generation scalar and vector LQs that can produce a pair of RHNs decaying to produce monoelectron and dielectron final states. Due to reduced backgrounds, the dielectron channel shows better prospects than the monoelectron mode. However, both channels show good promise at the HL-LHC. If the LQ-RHN-quark coupling is large, the HL-LHC can probe LQ masses between $3$--$10$ TeV (depending on the type of the LQ) for a sub-TeV RHN. In the future, machine learning techniques can enhance the prospects even further. While the model-independent mass limits derived from the pair production of vector LQs depend on the choice of the $\kappa$ parameter, the model-dependent limits are only marginally
affected by $\kappa$, as the t-channel LQ exchange process essentially determines them as long as the new coupling controlling it is not very small.

\section*{acknowledgement}
\noindent T.M. acknowledges partial support from the SERB/ANRF, Government of India, through the Core Research Grant (CRG) No. CRG/2023/007031. R.S. acknowledges the PMRF (ID: 0802000) from the Government of India.

\bibliography{References}
\bibliographystyle{JHEPCust}

\end{document}